\documentclass[twocolumn]{aastex631}
\usepackage{graphicx}
\usepackage{amsmath}
\usepackage{amssymb}
\usepackage{supertabular}
\usepackage{hyperref}
\usepackage{xcolor}
\usepackage{academicons}
\usepackage{booktabs}
\usepackage[normalem]{ulem}  

\begin{document}
\title{Exploring nuclear force with pulsar glitch observation}

\author[0000-0001-6836-9339]{Zhonghao Tu}
\affiliation{Department of Astronomy, Xiamen University, Xiamen, Fujian 361005, China; tuzhonghao@xmu.edu.cn; liang@xmu.edu.cn}
\author[0000-0001-9849-3656]{Ang Li}
\affiliation{Department of Astronomy, Xiamen University, Xiamen, Fujian 361005, China; tuzhonghao@xmu.edu.cn; liang@xmu.edu.cn}

\date{\today}

\begin{abstract} 
We connect nuclear forces to one of the most notable irregular behaviors observed in pulsars, already detected in approximately 6\% known pulsars, with increasingly accurate data expected from upcoming high-precision timing instruments on both ground and space.
Built on \cite{Shang2021_ApJ923-108}, we conduct a case study on the 2000 glitch of the Vela pulsar.
For our purpose, we adopt the Relativistic Mean Field (RMF) model as the theoretical many-body framework to describe nuclear systems. We refit three representative RMF parameter sets (DD-ME2, PKDD, NL3), considering the uncertainties in nuclear matter saturation properties.
Utilizing the resulting star structure, composition and nucleon properties in the medium obtained in a consistent manner, we calculate the pinning energy of superfluid vortex in the nuclear lattice in the inner crust. This leads to the evolution of associated pinning force that acts on the vortex, which can be confronted with observed glitch amplitude and short-time relaxation in the 2000 Vela glitch event, following the snowplow model of pulsar glitch.
We discuss how the vortex configuration and pinning properties depend on the nuclear parameters, and find an interesting and dominant role of the nuclear symmetry energy slope on pinning strength.

\end{abstract}


\section{Introduction}
Now approximately 6$\%$ of known pulsars have been identified through long-term pulsar timing observations to exhibit the so-called glitches \citep{Espinoza2011_MNRAS414-1679,Yu2012_MNRAS429-688,Basu2021_MNRAS510-4049,Lower2021_MNRAS508-3251,Zubieta2024_AA689-A191}, i.e., sudden increases in the rotational frequency followed by a relaxation over days to years \citep{Haskell2015_IJMPD24-1530008,Antonopoulou2022_RPP85-126901,Zhou2022_Universe8-641,Antonelli2022_NSPhysicsGlitches}.
Specifically, a total of 25 glitches of the Vela pulsar have been detected over the past 56 years \citep{Basu2021_MNRAS510-4049}, and have provided unique insights into the internal structure of neutron stars (NSs) and the equation of state (EoS) of dense matter \citep{Baym1969_Nature224-872,Haskell2013_ApJ764-L25,Alpar1993_ApJ409-345,Li2016_ApJSupp223-16,Yan2019_RAA19-072,Guegercinoglu2020_MNRAS496-2506,Shang2021_ApJ923-108}. For Vela-like pulsars, i.g., pulsars with glitch behaviors similar to Vela, glitches are typically attributed to complicated vortex dynamics within the pulsar, which can be described by the vortex-mediated glitch scenario \citep{Baym1969_Nature224-872,Anderson1975_Nature256-25,Pizzochero2011_ApJ743-L20,Haskell2011_MNRAS420-658,Graber2018_ApJ865-23,Guegercinoglu2020_MNRAS496-2506}. Some laboratory experiments in superfluid helium \citep{Campbell1979_PRL43-1336,Tsakadze1980_JLTP39-649} and simulations in rotating supersolids \citep{Poli2023_PRL131-223401,Bland2024_FBS65-81,Alana2024_PRA110-023306} have provided support to this scenario.

The snowplow model~\citep{Pizzochero2011_ApJ743-L20,Seveso2012_MNRAS427-1089} describes the vortex dynamics within a star by incorporating the pinning forces throughout the stellar crust, based on which strong constraints on the EoS and the mass of Vela can be obtained \citep[e.g.,][]{Hooker2013_JPCS420-012153,Shang2021_ApJ923-108}.
The pinning force, which bridges the microscopic EoS and glitch observations, plays an important role in the snowplow model.
In previous works, the pinning force used in the snowplow model is typically represented in an analytic form \citep{Pizzochero2011_ApJ743-L20,Seveso2012_MNRAS427-1089,Shang2021_ApJ923-108}, where the density corresponding to the maximum value of the pinning force is determined by the polarization strength of the superfluid neutrons \citep{Grill2012_JPCS342-012004,Seveso2016_MNRAS455-3952}. However, a more proper evaluation of the pinning force can be calculated from the realistic pinning energy, which is related to the pairing properties of the neutrons and the detailed structure of the inner crust of NSs.

Furthermore, early calculations of pinning energy only considered the contribution from the pairing condensation energy of superfluid neutrons inside the vortex core \citep{Alpar1977_ApJ213-527,Alpar1984_ApJ276-325}. Based on the local density approximation (LDA), the semi-classical calculations have been developed by utilizing a more realistic density distribution of the inner crust at a given density and further incorporating the contribution of kinetic energy of superfluid neutrons outside the vortex core \citep{Negele1973_NPA207-298,Epstein1988_ApJ328-680,Pizzochero1997_PRL79-3347,Donati2004_NPA742-363}. 
The semi-classical calculations have shown that vortices are pinned to nuclei in the deeper layer of the inner crust, while are repelled at lower density.
Improved results can be obtained by including kinetic corrections related to the presence of the nucleus in the vortex flow and the quantum structure of the vortex core \citep{Donati2004_NPA742-363}.

More microscopically, the non-relativistic axially symmetric Hartree-Fock-Bogoliubov (HFB) calculations were developed by \cite{Avogadro2008_NPA811-378} and \cite{Klausner2023_PRC108-035808}. They found that the pinning energy strongly depends on both the nuclear interaction and polarization. For example, when SkM$^*$ is adopted, independent of polarization strength, vortices are pinned to the top of a nucleus at lower density, but are repelled by the nuclei at deeper layers of the inner crust.
For calculations with SLy4 and without polarization, the pinning properties are similar to those of SkM$^*$, but vortices can pin to nuclei in the deeper layer of the inner crust. In addition, the pinning strengths calculated with SLy4 are weaker than those calculated with SkM$^*$. The dependence of the pinning energy on nuclear interactions and polarization may significantly influence the dynamical properties of vortices in the inner crust, leading to observable effects in glitches.
However, up to now, such studies usually neglect the effect of realistic pinning energy, e.g., \cite{Hooker2013_JPCS420-012153} and \cite{Yan2019_RAA19-072}.

In this work, with the attempt to better understand the nuclear interaction and its pairing property with glitch observations, we adopt one model of quantum hadrodynamics \citep{Fetter1971_PT25-54,Walecka1974_APNY83-491,Serot1992_RPP55-1855}, e.g., relativistic mean field (RMF) model to construct the unified EoSs for NSs with different nuclear effective interactions and self-consistently calculated the composition of the inner crust with different polarization strengths. The neutron pairing gap is calculated separately by the standard Bardeen-Cooper-Schrieffer (BCS) approximation.
The pinning energy calculations are performed using the semi-classical method from which we obtain the pinning properties across the entire density range of the inner crust.

This paper is organized as follows. In Sec.~\ref{sec:theor_eos}--\ref{sec:theor_pinning}, the theoretical framework for constructing unified EoS, the composition of NSs, the neutron pairing gap, and the pinning energy are given, respectively.  Sec. \ref{sec:theor_snowplow} is dedicated to formulating the snowplow model. In Sec.~\ref{sec:res}, we demonstrate the dependence of pinning energy and pinning force on the effective interactions and polarization, and present the constraint on the nuclear parameters and Vela mass with the 2000 Vela glitch observation. Some necessary discussions are also given in Sec. \ref{sec:res}.  Finally, a brief summary is given in Sec.~\ref{sec:summary}.

\section{Theoretical Framework}\label{sec:theor}

In this section, we present the necessary formulas for describing many-body nucleonic systems, single-particle properties, and pinning characteristics.

\subsection{RMF framework for nuclear matter and refitting parameter sets}\label{sec:theor_eos}

For describing nuclear matter, we consider the baryons interacting with each other through the exchange of isoscalar scalar and vector mesons ($\sigma$ and $\omega$), isovector vector meson ($\rho$) in the RMF model.
The Lagrangian density that describes the systems with time-reversal symmetry can be written as \citep{Boguta1977_NPA292-413,Reinhard1989_RPP52-439,Fuchs1995_PRC52-3043,Reinhard1989_RPP52-439,Ring1996_PPNP37-193,Bender2003_RMP75-121,Niksic2011_PPNP66-519,Oertel2017_RMP89-015007}:
\begin{widetext}
\begin{equation}\label{equ:RMF_Lagrangian}
\begin{aligned}
    \mathcal{L} = & \sum_{N=n,p}\bar{\psi}_{N}\left\{ \gamma^{\mu}\left[ i\partial_{\mu}-g_{\omega N}\omega_{\mu}
                 -g_{\rho N}\boldsymbol{\rho}_{\mu}\boldsymbol{\tau}_{N}-q_NA_{\mu} \right]
                 -\left[ M_{N}-g_{\sigma N}\sigma\right] \right\}\psi_{N} \\
                &+\frac{1}{2}(\partial^{\mu}\sigma\partial_{\mu}\sigma-m_{\sigma}^{2}\sigma^{2})+U(\sigma,\omega_\mu,\boldsymbol{\rho}_{\mu})\\
                &-\frac{1}{4}W^{\mu\nu}W_{\mu\nu}+\frac{1}{2}m_{\omega}^{2}\omega^{\mu}\omega_{\mu}
                 -\frac{1}{4}\boldsymbol{R}^{\mu\nu}\boldsymbol{R}_{\mu\nu}+\frac{1}{2}m_{\rho}^{2}\boldsymbol{\rho}^{\mu}\boldsymbol{\rho}_{\mu}-\frac{1}{4}A^{\mu\nu}A_{\mu\nu}\\
                &+\sum_{l=e,\mu}\bar{\psi}_{l}(i\gamma_{\mu}\partial^{\mu}-m_{l}+e\gamma^0A_\mu)\psi_{l},
\end{aligned}
\end{equation}
\end{widetext}
where~$\boldsymbol{\tau}_{N}$ and $q_N$ are the Pauli matrices and charges of nucleons, and~$M_{N}$~and~$m_{l}$~represent the nucleon and lepton masses, respectively. $g_{mN}$ is the coupling constant between the meson $m$ and the nucleon $N$. $\psi_{N(l)}$~is the Dirac field of the nucleons or the leptons. $\sigma$, $\omega_{\mu}$, and $\boldsymbol{\rho}_{\mu}$ denote the quantum fields of mesons. The field tensors of $\omega$, $\rho$, and photon are
\begin{equation}\label{equ:fieldtensor}
\begin{aligned}
    W_{\mu\nu}       &= \partial_{\mu}\omega_{\nu}-\partial_{\nu}\omega_{\mu}, \\
    \boldsymbol{R}_{\mu\nu} &= \partial_{\mu}\vec{\rho}_{\nu}-\partial_{\nu}\vec{\rho}_{\mu}, \\
    A_{\mu\nu} &= \partial_{\mu}A_{\nu}-\partial_{\nu}A_{\mu}.
\end{aligned}
\end{equation}
Under the mean-field approximation, all meson fields are treated as classical fields and we replace them by the corresponding mean values. The equations of motion of various mesons can be obtained via the Euler-Lagrange equation
\begin{align}
    (-\nabla^2+m_{\sigma}^{2})\sigma &= \sum_{N}g_{\sigma N}\rho_{s}^{N}+\frac{\partial U}{\partial\sigma}, \label{equ:equ_motion_meson_1}\\
    (-\nabla^2+m_{\omega}^{2})\omega &= \sum_{N}g_{\omega N}\rho_{v}^{N}-\frac{\partial U}{\partial\omega},\label{equ:equ_motion_meson_2} \\
    (-\nabla^2+m_{\rho}^{2})\rho     &= \sum_{N}g_{\rho N}\rho_{v}^{N}\tau_{N}^{3}-\frac{\partial U}{\partial\rho},\label{equ:equ_motion_meson_3} \\
    -\nabla^2A_0 &= e(\rho_v^p-\rho_v^e-\rho_v^\mu), \label{equ:equ_motion_meson_4}
\end{align}
where $\tau_{N}^{3}$ is the isospin projection of nucleons, $\rho_v^N$ and $\rho_s^N$ are the vector and scalar densities, respectively,
\begin{equation}\label{equ:scalar+vector_density}
\begin{aligned}
    \rho_{v}^{N} &= \frac{1}{\pi^{2}}\int_{0}^{k_{\mathrm{F}}^{N}}k^{2}\mathrm{d}k = \frac{(k_{\mathrm{F}}^{N})^{3}}{3\pi^{2}}, \\
    \rho_{s}^{N} &= \frac{M_{N}^{\star}}{\pi^{2}}\int_{0}^{k_{\mathrm{F}}^{N}}\frac{k^{2}\mathrm{d}k}{\sqrt{k^{2}+M_{N}^{\star 2}}}\\
                   &= \frac{(M_{N}^{\star})^{3}}{2\pi^{2}}\left[ q\sqrt{1+q^{2}}-\ln(q+\sqrt{1+q^{2}}) \right],
                     \quad q=\frac{k_{\mathrm{F}}^{N}}{M_{N}^{\star}},
\end{aligned}
\end{equation}
with the Fermi momentum $k_{\mathrm{F}}^{N}$ and the Dirac effective mass $M_{N}^{\star}=M_{N}-g_{\sigma N}\sigma$.
Eqs. (\ref{equ:equ_motion_meson_1})-(\ref{equ:equ_motion_meson_3}) for the meson fields, coupled to the Dirac equations for the nucleons, are solved self-consistently in the RMF approximation. Then the nuclear contribution to the energy of the system can be evaluated (see immediately below in Sec. \ref{sec:theor_crust}).

The interaction term in the RMF Lagrangian Eq.~(\ref{equ:RMF_Lagrangian}) depends on the nucleon-meson coupling constants that are usually determined
by fitting nuclei or nuclear-matter properties.
For a quantitative description of nuclear matter and finite nuclei, one needs to include a medium dependence of the effective mean-field interactions accounting for higher-order many-body effects.
The medium dependence can either be introduced by including non-linear (NL) meson self-interaction terms in the Lagrangian, or by assuming an explicit density dependence (DD) for the meson-nucleon couplings.
In the NLRMF model, self-coupling terms and cross-coupling terms for the mesons are introduced in $U(\sigma,\omega_\mu,\boldsymbol{\rho}_{\mu})$ with the flexibility of adjusting the density dependence of the symmetry energy, defined based on the parabolic approximation of the binding energy per nucleon $E/A = \varepsilon/\rho- M_N$:
$E/A (\rho_{\rm B}, \beta) =  E/A (\rho_{\rm B}, \beta=0) + E_{\rm sym}(\rho_{\rm B})\beta^2 + ...$
with $\beta = (\rho_n-\rho_p)/\rho_{\rm B}$ denoting the isospin asymmetry.
Density-dependent coupling parameters in the DDRMF model correspond to
$U(\sigma,\omega_\mu,\boldsymbol{\rho}_{\mu})$ in the NLRMF model.
The density dependence of the coupling can be in phenomenological functional form, with parameters adjusted to experimental data. For example, following the Typel-Wolter ansatz \citep{Typel1999_NPA656-331}, the density dependence of the coupling parameters are
\begin{equation}\label{equ:DD_for_sigma_and_omega}
\begin{aligned}
    g_m(\rho_{\rm{B}})=g_m(\rho_0)a_m\frac{1+b_m(\rho_{\rm{B}}/\rho_0+d_m)^2}{1+c_m(\rho_{\rm{B}}/\rho_0+e_m)^2}
\end{aligned}
\end{equation}
for $m=\sigma$ or $\omega$ and
\begin{equation}\label{equ:DD_for_rho}
\begin{aligned}
    g_\rho(\rho_{\rm{B}})=g_\rho(\rho_0)\exp\left[-a_\rho(\rho_{\rm{B}}/\rho_0-1)\right],
\end{aligned}
\end{equation}
for $\rho$ meson. Here $\rho_0$ is the nuclear saturation density.

\begin{table}[htbp]
  \centering
  \caption{Saturation properties of nuclear matter for different RMF effective interactions. The saturation properties we list below include the saturation density~$\rho_{0}$~(fm$^{-3}$), binding energy per particle $E/A$ (MeV), incompressibility $K_0$ (MeV), skewness $Q_0$ (MeV), symmetry energy $J_0$ (MeV), slope of symmetry energy $L_0$ (MeV), and effective mass of neutron $M_{n}^{\star}/M_{n}$.}\label{tab:satpros}
  \setlength\tabcolsep{3pt}
  \begin{tabular}{lccccccc}
  \hline
  \hline
         & $\rho_{0}$ & $E/A$ & $K_0$ & $Q_0$ & $J_0$ & $L_0$ & $M_n^{\star}/M_n$ \\
                          & $\rm{fm}^{-3}$ & MeV & MeV & MeV & MeV & MeV \\
  \hline
    DD-ME2 & 0.152 & -16.14 & 251.1 & 479 & 32.30 & 51.26 & 0.572 \\
    PKDD   & 0.150 & -16.27 & 262.2 & -115 & 36.86 & 90.21 & 0.570 \\
    NL3    & 0.148 & -16.24 & 272.2 & 198 & 37.4 & 118.5 & 0.594 \\
  \hline
  \end{tabular}
\end{table}

\begin{table*}[htbp]
  \centering
  \caption{The coupling parameters $g_{\rho}$ and $a_{\rho}$ between nucleons and $\rho$ meson for symmetry energy slope $L_0=30, 40, 60, 80$ MeV. These parameters are obtained by fixing the symmetry energy $E_{\rm{sym}}$ at $\rho_{\mathrm{B}}=0.11$ fm$^{-3}$ but adjusting the symmetry energy slope $L_0$ at the saturation density.}\label{tab:vectorcouple}
  \setlength\tabcolsep{15pt}
  \begin{tabular}{lcccccccc}
    \hline
    \hline
    $L_0$ (MeV) & \multicolumn{2}{c}{30} & \multicolumn{2}{c}{40} & \multicolumn{2}{c}{60} & \multicolumn{2}{c}{80} \\
    \cmidrule(r){2-3} \cmidrule(r){4-5} \cmidrule(r){6-7} \cmidrule(r){8-9}
      & $g_{\rho}$ & $a_{\rho}$ & $g_{\rho}$ & $a_{\rho}$ & $g_{\rho}$ & $a_{\rho}$ & $g_{\rho}$ & $a_{\rho}$ \\
    \hline
    DD-ME2 & 3.3574 & 0.8914 & 3.5283 & 0.7207 & 3.7917 & 0.4599 & 4.0097 & 0.2576 \\
    PKDD   & 3.6289 & 0.8273 & 3.7686 & 0.6839 & 4.0055 & 0.4525 &  4.2063 & 0.2669 \\
    NL3    & -- & -- & 3.6901 & 0.7471 & 3.9359 & 0.4971 & 4.3241 & 0.1323 \\
    \hline
  \end{tabular}
\end{table*}

The RMF model provides an excellent tool to study the properties of finite nuclei and infinite nuclear matter \citep{Walecka1974_APNY83-491,Glendenning1996_CompactStars,Shen2002_PRC65-035802,Dutra2014_PRC90-055203,2020JHEAp..28...19L,Rong2021_PRC104-054321,Tu2022_ApJ925-16,Sun2023_ApJ942-55,2023PhRvC.108b5809Z}.
See \cite{Dutra2014_PRC90-055203} for a recent comparison of different RMF parameterizations.
The extensions of several effective interactions, e.g., DD-ME2 \citep{Lalazissis2005_PRC71-024312,Li2019_PRC100-015809}, PKDD \citep{Long2004_PRC69-034319}, and NL3 \citep{Schaffner-Bielich2000_PRC62-034311,Wu2021_PRC104-015802}, can be utilized to study the impact of nuclear symmetry energy on the pinning properties, with different density-slope of the symmetry energy at saturation density, $L_0=3\rho_0 (\frac{\mathrm{d}E_{\rm sym}}{\mathrm{d}\rho_{\rm B}})|_{\rho_0}$. Except for pulsar glitch, the other NS phenomena, e.g., cooling \citep{Newton2013_ApJL779-L4} and asteroseismology \citep{Neill2024_MNRAS532-827}, have been used to constrain $L_0$.

Table \ref{tab:satpros} lists the characteristic coefficients of nuclear matter for the original effective interactions DD-ME2, PKDD, and NL3. The original DD-ME2 is a widely-adopted interaction that effectively describes the properties of finite nuclei and nuclear matter, having the smallest symmetry energy slope $L_0=$ 51.26 MeV among the three effective interactions. In contrast, the original NL3 is characterized by an excessively large symmetry energy slope of $L_0=$ 118.5 MeV.

By adjusting the interaction in the isovector channel, we can revisit the compatibility of these effective interactions with observations. For DD-ME2 and PKDD, we typically fix the symmetry energies at the sub-saturation density $\rho_{\mathrm{B}}=0.11$ fm$^{-3}$ to their original values and the symmetry energy slopes at the corresponding saturation densities to our desired values, this allows us to determine the coupling coefficients $a_{\rho}$ and $g_{\rho}(\rho_0)$. For nonlinear effective interactions, the symmetry energy slope is usually adjusted by altering the $\omega$-$\rho$ coupling, as in the case of NL3$\omega\rho$ model; in this work, we adopt an alternative equivalent approach by introducing density-dependent coupling for the $\rho$ meson, with the density dependence of the coupling parameters described by Eq. (\ref{equ:DD_for_rho}). \footnote{Unlike \citet{Wu2021_PRC104-015802}, we fix the symmetry energy at sub-saturation densities as we do for DD-ME2 and PKDD.}
Therefore, in our work, the interactions in the isovector channel of all three effective interactions have been adjusted by using a fairly consistent approach.

In the following text, we use the names of the original effective interactions to refer to the interactions in the isoscale channel, whereas the corresponding symmetry energy slopes are used to refer to the interactions in the isovector channel.

Table \ref{tab:vectorcouple} presents the coupling parameters $g_{\rho}(\rho_0)$ and $a_{\rho}$ for different effective interactions with $L_0=$30, 40, 60, and 80 MeV. NL3 with $L_0=$ 30 MeV is excluded from our work due to its non-monotonic behavior in the EoS.

\subsection{Unified description of stellar crust and core}\label{sec:theor_crust}

We 
proceed to utilize the RMF model to construct crucial physical inputs for simulating glitches: the unified EoSs and self-consistent compositions of NSs.
Application of the unified EoS is beneficial to reduce the additional degrees of freedom introduced by the manual matching of the crust and core EoSs.

For uniform nuclear matter in the core, due to the translational invariance of the system, all derivative terms in Eqs. (\ref{equ:equ_motion_meson_1})--(\ref{equ:equ_motion_meson_4}) vanish. Using the Thomas-Fermi approximation, the nucleon Dirac field can be considered as plane wave, leading to a nucleon single-particle energy,
\begin{equation}\label{equ:sp_energy}
\begin{aligned}
  E_{N}(k) &=\sqrt{k^{2}+M_{N}^{*2}}+g_{\omega N}\omega
    +g_{\rho N}\tau_{N}^{3}\rho+g_{\phi N}\phi+\Sigma_{R}.
\end{aligned}
\end{equation}
The value of $E_{N}(k)$ at the Fermi surface is the corresponding chemical potential. $\Sigma_{R}$ is the rearrangement term that originated from the density dependence of the coupling constants between nucleons and mesons, $\Sigma_{R}=0$ for the NLRMF model. By combining the $\beta$-equilibrium condition, the charge neutrality condition, and the nucleon number conservation condition for uniform nuclear matter in the core, the nucleon Fermi momentum and meson fields at a given number density can be self-consistently solved.

Furthermore, the energy density $\varepsilon$ and pressure $P$ can be calculated by using the energy–momentum tensor,
\begin{equation}\label{equ:RMF_EoS}
\begin{aligned}
    \varepsilon = & \frac{1}{2}m_{\sigma}^{2}\sigma^{2}+\frac{1}{2}m_{\omega}^{2}\omega^{2}+\frac{1}{2}m_{\rho}^{2}\rho^{2}-U(\sigma,\omega,\rho)\\
                    &+\frac{\partial U}{\partial\omega}\omega+\frac{\partial U}{\partial\rho}\rho+\sum_{N}\varepsilon_{\mathrm{kin}}^{N}+\sum_{l}\varepsilon_{\mathrm{kin}}^{l}, \\
        P       = & -\frac{1}{2}m_{\sigma}^{2}\sigma^{2}+\frac{1}{2}m_{\omega}^{2}\omega^{2}+\frac{1}{2}m_{\rho}^{2}\rho^{2}
        +\rho\Sigma_{N}^{r}\\
        &+U(\sigma,\omega,\rho)+\sum_{N}P_{\mathrm{kin}}^{N}+\sum_{l}P_{\mathrm{kin}}^{l},
\end{aligned}
\end{equation}
where $\Sigma_R=0$ for NLRMF model and $U(\sigma,\omega,\rho)=0$ for DDRMF model. $\varepsilon_{\mathrm{kin}}^{N}$ ($\varepsilon_{\mathrm{kin}}^{l}$) and $P_{\mathrm{kin}}^{N}$ ($P_{\mathrm{kin}}^{l}$) are the contributions from kinetic energy,
\begin{equation}\label{equ:RMF_EoS_kin}
\begin{aligned}
    \varepsilon_{\mathrm{kin}} &= \frac{k_{\mathrm{F}}^{4}}{\pi^{2}}\left[ \left(1+\frac{z^{2}}{2}\right)\frac{\sqrt{1+z^{2}}}{4}
                                   -\frac{z^{4}}{8}\ln\left(\frac{1+\sqrt{1+z^{2}}}{z}\right) \right], \\
          P_{\mathrm{kin}}     &= \frac{k_{\mathrm{F}}^{4}}{3\pi^{2}}\left[ \left(1-\frac{3z^{2}}{2}\right)\frac{\sqrt{1+z^{2}}}{4}
                                   +\frac{3z^{4}}{8}\ln\left(\frac{1+\sqrt{1+z^{2}}}{z}\right) \right],
\end{aligned}
\end{equation}
where $z=1/q$.

\begin{figure}
\centering
\includegraphics[width=0.45\textwidth]{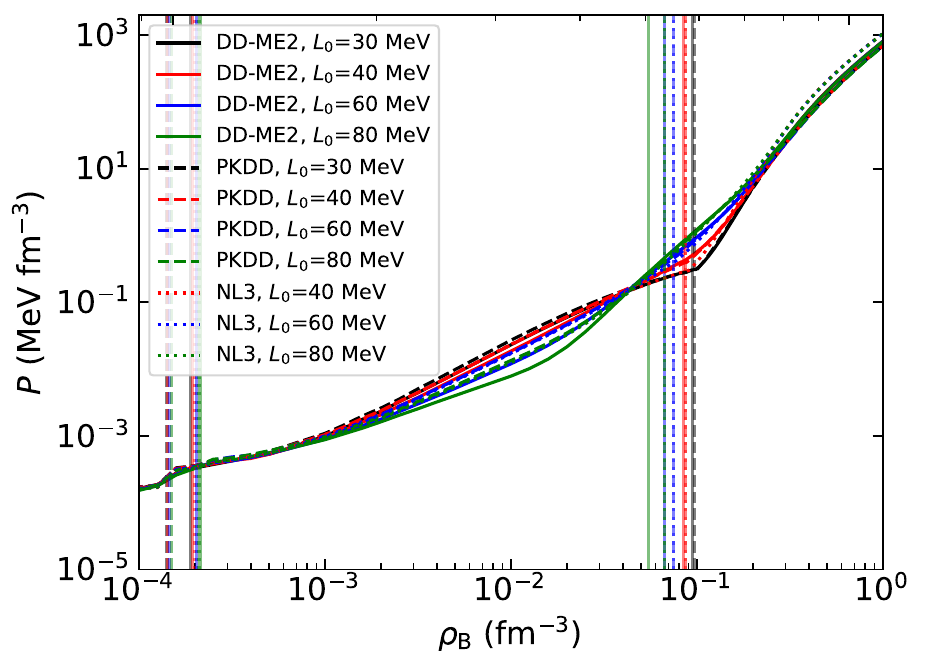}
\caption{Unified EoSs calculated with DD-ME2 (solid lines), PKDD (dashed lines), and NL3 (dotted lines) with the different symmetry energy slope. The vertical line with the same line style as the EoS represents the corresponding crust-core transition density.
}
\label{fig:EoS}
\end{figure}

\begin{table*}
  \centering
  \caption{The transition density of EoS and global properties of NSs for different RMF effective interactions. The transition density of EoSs we list below include the outer-inner crust transition density $\rho_{\rm{oi}}$ and the crust-core transition density $\rho_{\rm{cc}}$. We also list the maximum mass $M_{\rm{TOV}}$ of NSs for different effective interactions.
  }\label{tab:EoSproperties}
  \setlength\tabcolsep{9pt}
  \begin{tabular}{lccccccccccc}
    \hline
    \hline
     & \multicolumn{4}{c}{DD-ME2} & \multicolumn{4}{c}{PKDD} & \multicolumn{3}{c}{NL3} \\
     \cmidrule(r){2-5} \cmidrule(r){6-9} \cmidrule(r){10-12}
     $L_0$ (MeV) & 30 & 40 & 60 & 80 & 30 & 40 & 60 & 80 & 40 & 60 & 80 \\
    \hline
    $\rho_{\rm{oi}}$ ($10^{-4}$ fm$^{-3}$) & 1.902 & 1.944 & 2.035 & 2.129 & 1.416 & 1.431 & 1.463 & 1.496 & 1.982 & 2.039 & 2.102 \\
    $\rho_{\rm{cc}}$ (fm$^{-3}$) & 0.095 & 0.085 & 0.067 & 0.055 & 0.097 & 0.087 & 0.075 & 0.067 & 0.087 & 0.075 & 0.067 \\
    $M_{\rm{TOV}}$ ($M_{\odot}$) & 2.497 & 2.491 & 2.477 & 2.468 & 2.356 & 2.350 & 2.336 & 2.325 & 2.759 & 2.746 & 2.739 \\
    \hline
  \end{tabular}
\end{table*}
\begin{figure*}[t]
\centering
\includegraphics[width=0.9\textwidth]{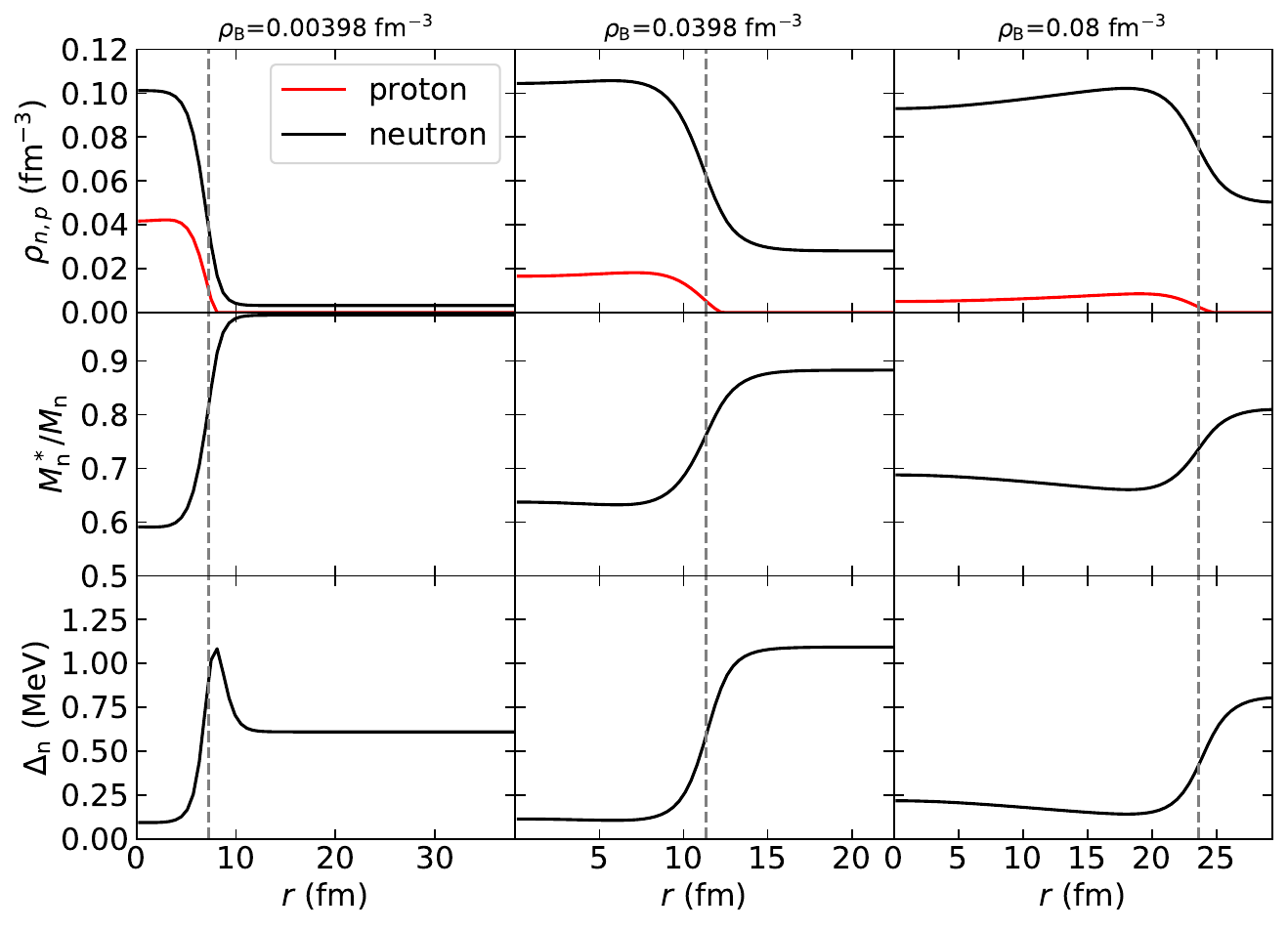}
\caption{The neutron and proton densities (upper panels), neutron $^1S_0$ pairing gap (middle panels), and neutron effective mass (lower panels) distributions within the WS cells at $\rho_{\mathrm{B}}$= 0.00398 fm$^{-3}$ (left panels), 0.0398 fm$^{-3}$ (middle panels), and 0.08 fm$^{-3}$ (right panels). The effective interaction is DD-ME2 with $L_0=$ 30 MeV, $\beta=3.0$. The gray dashed lines indicate the sizes of the droplet.}
\label{fig:density_gap_profile}
\end{figure*}

For the non-uniform matter of the outer and inner crusts, the microscopic nuclear structure can be described by the  Wigner-Seitz (WS) cell with different geometric parameters. The energy density $\varepsilon_{\rm{non}}$ is calculated by integrating the whole WS cell,
\begin{equation}\label{equ:energy_density_WS}
\begin{aligned}
  \varepsilon_{\rm{non}} &= \frac{\rho_{\mathrm{B}}}{A}\int_{0}^{R_{\rm{WS}}}4\pi\varepsilon r^2\rm{d}r,
\end{aligned}
\end{equation}
where $r$ is the distance of from the center of WS cell, $R_{\rm{WS}}$ is the radius of WS cell, and $A$ is the total number of nucleon within a WS cell. The local energy density is determined by the specific matter (mesons and fermions) distributions, see Sec. \ref{sec:theor_composition}. The pressure of non-uniform matter is then obtained:
\begin{equation}\label{equ:pressure_WS}
\begin{aligned}
  P_{\rm{non}} &= \sum_{i=N,l}\mu_iN_i/V-\varepsilon_{\rm{non}}.
\end{aligned}
\end{equation}
$N_i$ is the total number of nucleons and leptons, $V=4\pi R_{\rm{WS}}^3/3$ is the volume of WS cell.

Apply the spherical WS approximation and reflective boundary condition, the distributions of meson fields and densities of nucleons and leptons are obtained by iteratively solving the nonlinear system of equations containing Eqs. (\ref{equ:equ_motion_meson_1}-\ref{equ:equ_motion_meson_4}), $\beta$-equilibrium, charge neutrality, and nucleon number conservation conditions in the spherical symmetry one-dimensional geometry. In each iteration, the meson field is expanded by fast cosine transformation to improve the computational efficiency. The non-uniform structure is stable after fulfilling the constancy of chemical potentials,
\begin{equation}\label{equ:chem_cons_np}
\begin{aligned}
  \mu_N(r) =&\sqrt{k_{\rm{F}}^{N,2}(r)+M_{N}^{*2}(r)}+g_{\omega N}\omega(r)
    +g_{\rho N}\tau_{N}^{3}\rho(r)\\
    &+q_NA_0(r)+\Sigma_{R}(r) = \rm{constant},
\end{aligned}
\end{equation}
for nucleons $N=n,p$ and
\begin{equation}\label{equ:chem_cons_lepton}
\begin{aligned}
  \mu_l(r) =&\sqrt{k_{\rm{F}}^{l,2}(r)+m_{l}^{2}(r)}=\rm{constant},
\end{aligned}
\end{equation}
for leptons $l=e,\mu$. Here, $r$ again is the distance from the center of WS cell. The local energy density $\varepsilon$ at $r$ reads:
\begin{equation}\label{equ:energy_density_local}
\begin{aligned}
    \varepsilon = & \frac{1}{2}(\nabla\sigma)^2+\frac{1}{2}m_{\sigma}^{2}\sigma^{2}
    +\frac{1}{2}(\nabla\omega)^2\frac{1}{2}+\frac{1}{2}m_{\omega}^{2}\omega^{2}\\
    &+\frac{1}{2}(\nabla\rho)^2\frac{1}{2}+\frac{1}{2}m_{\rho}^{2}\rho^{2}
    +\frac{1}{2}(\nabla A_0)^2\frac{1}{2}-U(\sigma,\omega,\rho)\\
                    &+\frac{\partial U}{\partial\omega}\omega+\frac{\partial U}{\partial\rho}\rho+\sum_{N}\varepsilon_{\mathrm{kin}}^{N}+\sum_{l}\varepsilon_{\mathrm{kin}}^{l}.
\end{aligned}
\end{equation}

Fig. \ref{fig:EoS} shows the unified EoSs calculated with three sets of adopted effective interactions. In the intermediate densities, we observe significant differences in EoSs for different $L_0$; near the crust-core transition density $\rho_{\rm{cc}}$, a larger $L_0$ results in a stiffer EoS. As $g_{\rho}$ decreases with increasing density and approaches zero in the high density, the differences in the interactions of isovector channel diminish, then the EoSs are primarily controlled by the interactions in the isoscalar channel. NL3 yields the stiffest EoS at high densities, while PKDD gives the softest one.

The vertical lines on the left and right sides of Fig. \ref{fig:EoS} represent the outer-inner crust transition densities $\rho_{\rm{oi}}$ and crust-core transition densities $\rho_{\rm{cc}}$ derived from different effective interactions, respectively. Table \ref{tab:EoSproperties} lists their exact values.
The outer-inner crust transition density $\rho_{\rm{oi}}$ is fixed by the chemical potential of neutron exceeding its rest mass $M_N$, while the crust-core transition density $\rho_{\rm{cc}}$ is determined by $\varepsilon_{\rm{uni}}$ being less than  $\varepsilon_{\rm{non}}$.
It is clearly that as $L_0$ increases, $\rho_{\rm{oi}}$ increases while $\rho_{\rm{cc}}$ decreases, consistent with previous works \citep{Fantina2016_PRC93-015801}.

\subsection{Global properties and composition of neutron stars}
\label{sec:theor_composition}

The neutron vortices are pinned at the nuclear lattice of the inner crust of NSs, the properties of WS cell are crucial for the pinning.
The droplet size is important for the reasonable estimation of the kinetic energy contribution when a nucleus is fixed at the edge of a vortex. The droplet size is usually outlined by the density distribution of protons,
\begin{equation}\label{equ:droplet_size}
\begin{aligned}
  R_{\rm{d}}=R_{\rm{WS}}\left(\frac{\langle\rho_p\rangle^2}{\langle\rho_p^2\rangle}\right)^{1/3},
\end{aligned}
\end{equation}
with $\langle\rho_p\rangle=4\pi\int_{0}^{R_{\mathrm{WS}}}\rho_p r^2\mathrm{d}r/V$ and $\langle\rho_p^2\rangle=4\pi\int_{0}^{R_{\mathrm{WS}}}\rho_p^2 r^2\mathrm{d}r/V$.
The reliable method to calculate pinning energy that pinning to non-spherical nuclei is still lack. We reluctantly neglect the strange pasta structures, e.g., rod, tube, slab, bubble, in the bottom of the inner crust.

From $\rho_{\rm{oi}}$ to $\rho_{\rm{cc}}$, the properties of WS cell exhibit similar characteristics among all effective interactions. Taking DD-ME2 with $L_0=30$ MeV and $\beta=3.0$ as an example, Fig. \ref{fig:density_gap_profile} displays the nucleon density, neutron effective mass, and neutron pairing gap distributions within the WS cells at $\rho_{\mathrm{B}}=$0.00398, 0.0398, and 0.08 fm$^{-3}$. The gray vertical lines indicate the droplet sizes calculated by Eq. (\ref{equ:droplet_size}).
We also see that the relative volume of the droplet occupying the WS cell increases with increasing $\rho_{\mathrm{B}}$. In the left and middle panels of Fig. \ref{fig:density_gap_profile}, free neutrons occupy a large volume within the WS cell, resembling droplets bathed in the neutron sea. However, as $\rho_{\mathrm{B}}$ approaches $\rho_{\rm{cc}}$, the volume occupied by free neutrons decreases distinctly and the uniformity of the neutron sea is broken, resulting in the free neutrons moving through the gaps between tightly packed droplets.

\begin{figure}
\centering
\includegraphics[width=0.45\textwidth]{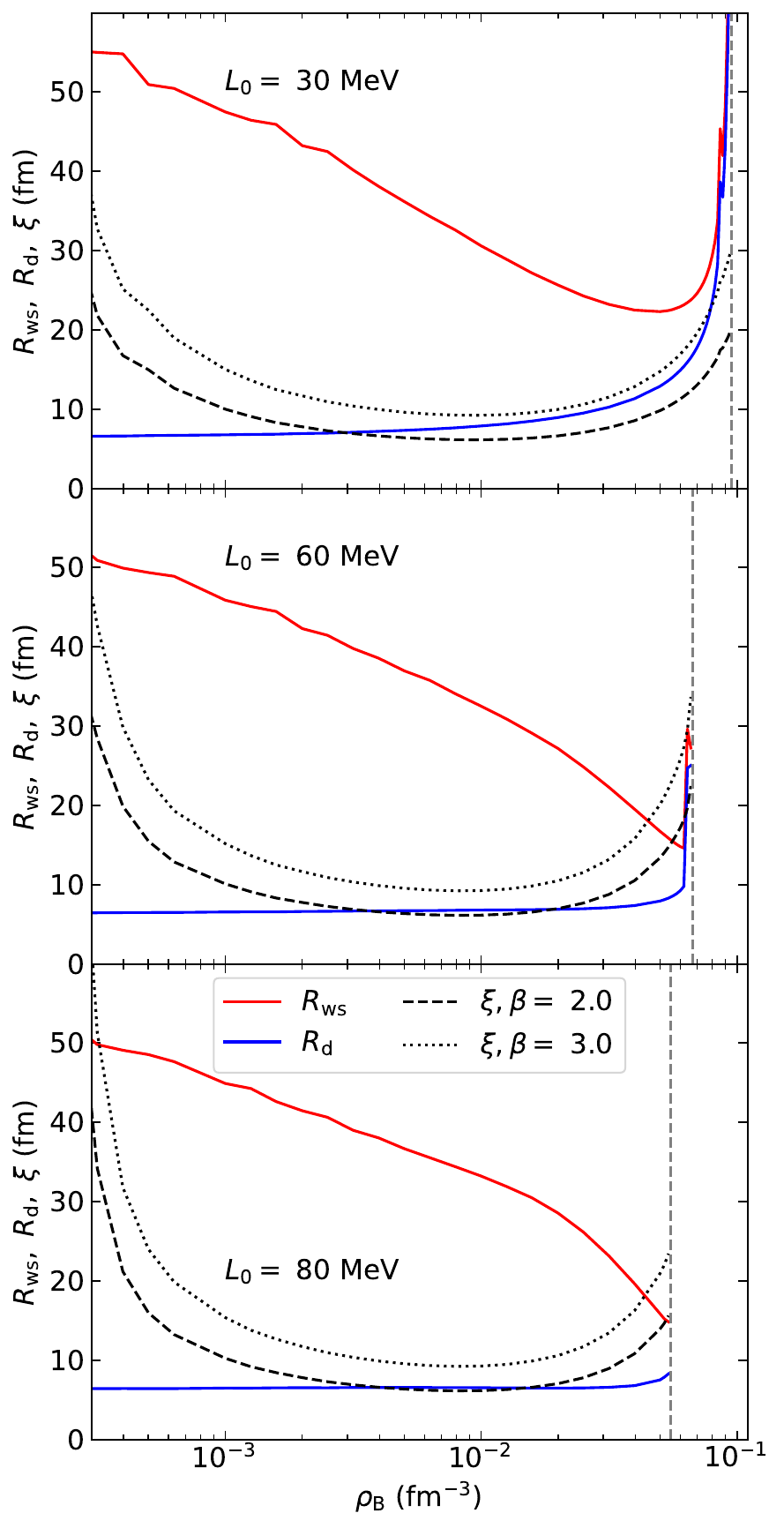}
\caption{The Wigner-Seitz cell size $R_{\rm{ws}}$ (red solid lines), droplet size $R_{\rm{d}}$ (blue solid lines), and superfluid coherence length $\xi$ (black dashed lines for $\beta$=2.0 and black dotted lines for $\beta$=3.0) for the inner crust with $L_0=$ 30, 60, and 80 MeV. DD-ME2 is adopted for the isoscalar channel of the effective interaction. The core-crust transition densities are depicted by gray dashed lines.}
\label{fig:size}
\end{figure}
\begin{figure*}
\centering
\includegraphics[width=0.90\textwidth]{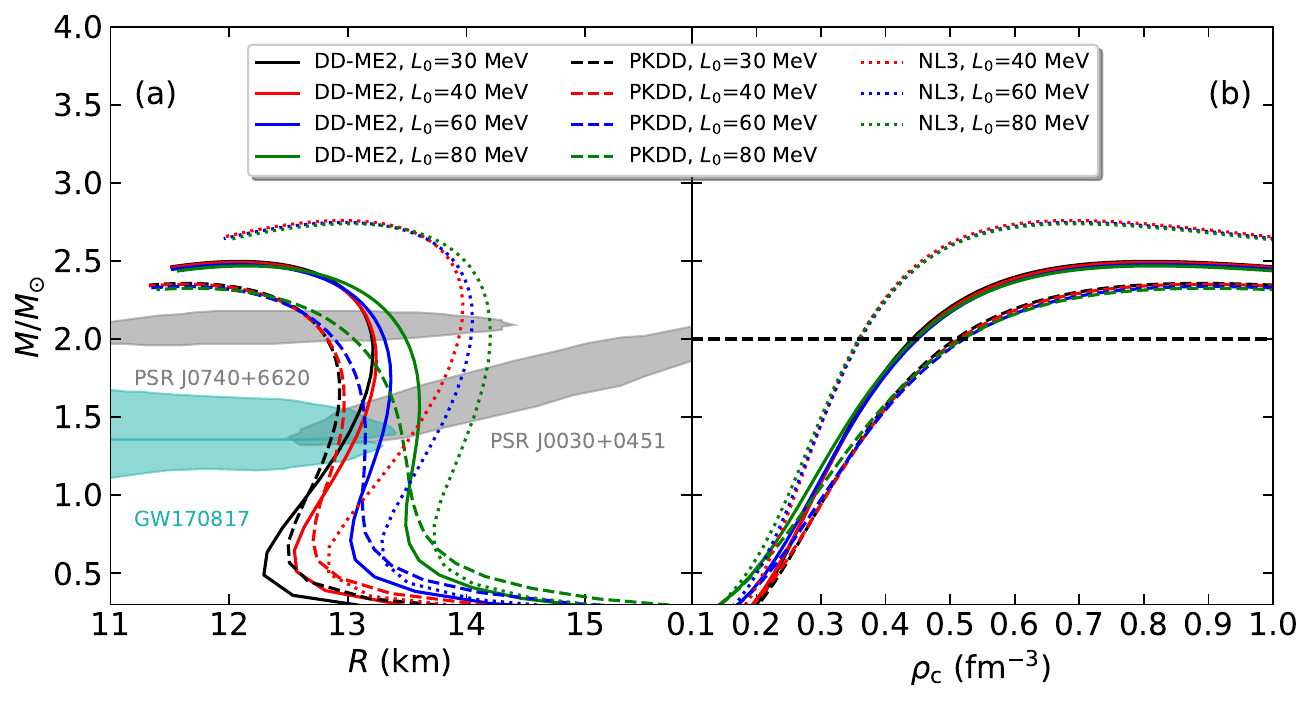}
\caption{(a) $M$-$R$ relations and (b) mass profiles calculated with the unified EoSs of DD-ME2 (solid lines), PKDD (dashed lines), and NL3 (dotted lines) with the different symmetry energy slope $L_0=$30, 40, 60, and 80 MeV. The mass-radius measurements from GW observations for GW170817 \citep{Abbott2017_PRL119-161101} and NICER missions for PSR J0030+0451 \citep{Vinciguerra2024_ApJ961-62} and PSR J0740+6620 \citep{Riley2021_ApJL918-L27} are shaded in panel (a). The dashed horizontal line presents $2M_{\odot}$ in panel (b).
}
\label{fig:MR_rhoc}
\end{figure*}
Now we discuss the dependence of droplet size $R_{\rm{d}}$ and WS cell size $R_{\rm{WS}}$ on effective interactions, particularly $L_0$. Fig. \ref{fig:size} illustrates $R_{\rm{d}}$ and $R_{\rm{WS}}$ calculated using DD-ME2 with $L_0=$30, 60, and 80 MeV as a function of $\rho_{\mathrm{B}}$. At lower densities, $L_0$ has a minor effect on $R_{\rm{d}}$ and $R_{\rm{WS}}$. However, near the crust-core transition densities, we can observe a dramatic increase in both $R_{\rm{d}}$ and $R_{\rm{WS}}$; this effect becomes more pronounced with smaller $L_0$ and tends to disappear at $L_0=$ 80 MeV. The effects of symmetry energy on $R_{\rm{d}}$ and $R_{\rm{WS}}$ can be explained by the surface tension. \citet{Oyamatsu2007_PRC75-015801} has showed that the surface tension decreases with increasing $L_0$. Consequently, droplets calculated at lower $L_0$ can maintain inhomogeneous structures under higher density and stronger Coulomb force due to the larger surface tension, thus further supporting a higher transition density and a larger droplet size. The upper panel of Fig. \ref{fig:size} for $L_0=$ 30 MeV suggests the possible existence of the giant droplet and extremely dense inhomogeneous structure near the crust-core transition density, which could significantly reduce pinning force.

Both the global properties and composition of NSs are essential to simulations of pulsar glitches.
Using the unified EoS shown in Fig.~\ref{fig:EoS} as basic input, the global properties of NSs, e.g., mass-radius ($M$-$R$) relation and mass profile, can be obtained by solving the Tolman--Oppenheimer--Volkoff (TOV) equation~\citep{Tolman1939_PR055-364,Oppenheimer1939_PR055-374},
\begin{equation}\label{equ:TOV}
\begin{aligned}
    \frac{\mathrm{d}P}{\mathrm{d}r} &= -\frac{\left[ P(r)+\varepsilon(r) \right]\left[ m(r)+4\pi r^{3}P(r) \right]}{r\left[ r-2m(r) \right]}, \\
    \frac{\mathrm{d}m}{\mathrm{d}r} &= 4\pi r^{2}\varepsilon(r),
\end{aligned}
\end{equation}
where $r$ is the distance from the center of NSs. Given a central density $\rho_{c}$, the TOV equation is integrated from $r=0$ to $r=R$ where the pressure is zero. $R$ is defined as the radius of NSs and $M=M(R)$ is the gravitational mass. We have taken $G=c=1$ in Eq. (\ref{equ:TOV}). The TOV equation also gives us the mass density profile $\rho(r)$ of an NS; note that in the following text $\rho$ denotes mass density not $\rho$ meson field. Then the moment of inertia $I(r_1,r_2)$ can be calculated under rigid-body rotation,
\begin{equation}\label{equ:MOI}
\begin{aligned}
I(r_1,r_2)=\frac{8\pi}{3}\int_{r_1}^{r_2}r^4\rho(r)\mathrm{d}r.
\end{aligned}
\end{equation}
The total moment of inertia is $I(0,R)$. The superfluid moment of inertia is obtained by simply replacing $\rho$ with $(1-Y_{p})\rho$ in Eq. (\ref{equ:MOI}), where $Y_{p}$ is the proton fraction for a given EoS.

Fig. \ref{fig:MR_rhoc} shows the $M$-$R$ relations and mass profiles calculated from different EoSs. As shown in Fig. \ref{fig:MR_rhoc}(a), all EoSs satisfy current observations of NS mass and radius. The maximum NS mass $M_{\rm{TOV}}$ corresponding to each EoS can be found in Table \ref{tab:EoSproperties}. The interaction in the isovector channel affects the radius of typical NSs ($\sim1.4M_{\odot}$): the larger $L_0$, the larger the radius. In Fig. \ref{fig:MR_rhoc}(b), we can see that the relations between central density of NSs $\rho_{\rm{c}}$ and NS masses are primarily determined by the interaction in the isoscalar channel for those NSs which are heavier than the typical NS.

\subsection{Neutron Pairing Gap}\label{sec:theor_gap}

Based on LDA, we calculate local neutron pairing gap within a WS cell. In the inner crust, neutrons form a $^1S_0$ superfluid. Using the BCS approximation \citep{Khodel1996_NPA598-390,Balberg1998_PRC57-409,Wang2010_PRC81-025801} and the phenomenological pairing force, the $^{1}S_0{}$ neutron pairing gap $\Delta(k)$ at zero temperature is obtained by solving the gap equation:
\begin{equation}\label{equ:gap_equ}
\begin{aligned}
    1 & = -\frac{1}{4\pi^{2}}\int k^{2}\mathrm{d}k\frac{G_{N}p^{2}(k)}{\sqrt{[E_{N}(k)-\mu]^{2}+\Delta_{0}^{2}p^{2}(k)}},
\end{aligned}
\end{equation}
where $E(k)$ is the neutron single particle energy obtained by RMF model, the chemical potential $\mu$ is determined by the value of $E(k)$ at the Fermi surface, i.e., $\mu = E(k_{\mathrm{F}})$ with the Fermi momentum $k_{\mathrm{F}}$.
$\Delta(k) = \Delta_0 p(k)$ is the trivial solution of the Eq.~(\ref{equ:gap_equ}) and we mainly refer the neutron pairing gap at the Fermi surface $\Delta_n\equiv\Delta_0 p(k_{\mathrm{F}})$.
In this work,
we adopt $G_{N}=738~\mathrm{MeV\cdot fm}^3$ for the neutron-neutron pairing strength, and a momentum dependence force $p(k)=\exp(-\alpha^2k^2)$ with a finite range of $\alpha = 0.636~\mathrm{fm}$, following \cite{Tian2009_PLB676-44} and \cite{Rong2020_PLB807-135533}.

From Eq. (\ref{equ:gap_equ}), the pairing gap is determined by $k_{\mathrm{F}}$ and $M_{n}^{\star}$. However, under the RMF model we are discussing, $M^{\star}_{n}$ are not affected by the isovector channel. Therefore, the pairing gap of pure neutron matter is strictly independent of $L_0$.
It is evident from Fig. \ref{fig:density_gap_profile} that as $\rho_{\mathrm{B}}$ increases, the density of the neutron sea $\rho_{\infty}$ (or the Fermi momentum of free neutrons $k_{\mathrm{F},\infty}$) increases while the effective mass of free neutrons $M_{\infty}^{\star}$ decreases. The subscript $\infty$ indicates the corresponding value when the density distribution is extrapolated to infinity. The neutron pairing gap is positively correlated with its Fermi momentum and effective mass; therefore, the pairing gap of free neutrons $\Delta_{\infty}$ increases at lower densities due to the significant increase in $k_{\mathrm{F},\infty}$, but decreases at higher densities due to the substantial reduction in $M_{\infty}^{\star}$.

Using the BCS approximation, Eq. (\ref{equ:gap_equ}) removes all contributions beyond the bare nuclear potential in the interaction kernel. In fact, medium polarization, namely the screening of nuclear medium on pairing, leads to an overall suppression to the pairing gap. The reduction of the pairing gap due to medium polarization can be simply represented by the polarization strength $\beta$ \citep{Donati2006_PLB640-74}, $\Delta_n(\beta)=\Delta_n/\beta$. \citet{Lombardo2000_arXiv0012209} have found that $\beta$ is between 2.0 and 3.0. In this work, we consider only the two cases of the weak polarization $\beta=2.0$ and strong polarization $\beta=3.0$.

\subsection{Nuclear Pinning Force}\label{sec:theor_pinning}

\begin{figure}
\centering
\includegraphics[width=0.45\textwidth]{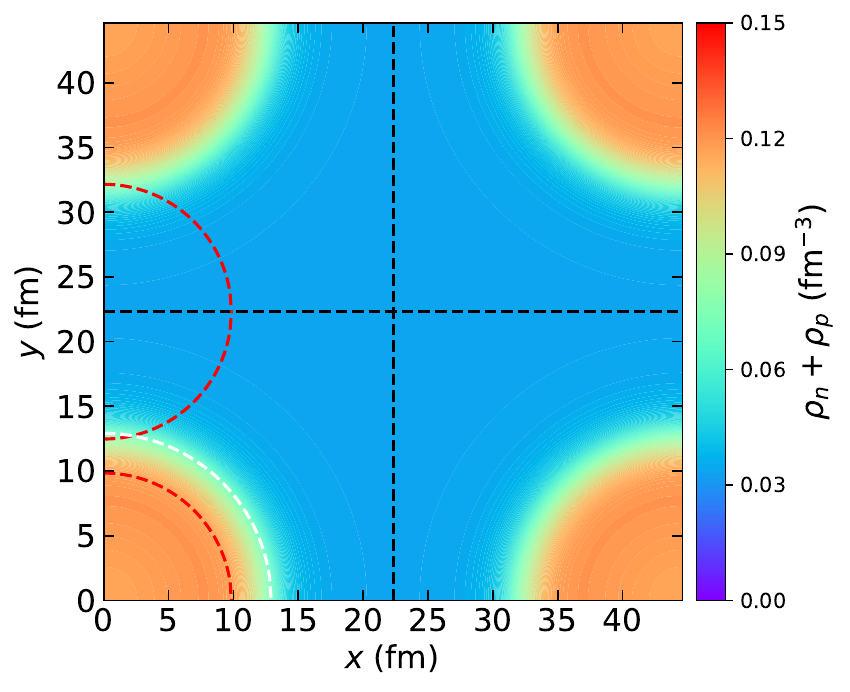}
\caption{The density profile of a cubic lattice in the xy plane for $\rho_{\mathrm{B}}=0.05$ fm$^{-3}$ and $z=0$.  The effective interaction DD-ME2 with $L_0=$ 30 MeV is adopted. The black dashed lines represent the boundary of the WS cell. The red dashed contours indicate the position of the vortex cores associated with nuclear pinning and interstitial pinning. The white dashed contour stands for the size of the droplet outlined by the proton density.}
\label{fig:cubic}
\end{figure}

The glitch dynamics is determined by pinning properties of neutron vortex in a nuclei lattice. A representative configuration in the inner crust of a NS is illustrated in Fig. \ref{fig:cubic}.
We will follow the general definition of pinning energy $E_{\rm{p}}$, which is the difference in the energy cost of vortices in the interstitial pinning (IP) and nuclear pinning (NP) configurations. Here, the NP configuration refers to the pinning of vortices on nuclei or droplets surrounded by the neutron sea, whereas the IP configuration refers to the pinning of vortices in the neutron sea between two adjacent nuclei.

The neutrons within a vortex are in a velocity field,
\begin{equation}\label{equ:neutron_velocity}
\begin{aligned}
  \boldsymbol{v(x)}=\frac{\hbar}{2M_{n}^{\star}r}\boldsymbol{e}_{\vartheta},
\end{aligned}
\end{equation}
where $r$ is the distance of point $\boldsymbol{x}$ within this vortex to the vortex axis. $\boldsymbol{e}_{\vartheta}$ is the tangent unit vector. Eq. (\ref{equ:neutron_velocity}) indicates that as the neutrons get closer to the vortex axis, their velocity and kinetic energy increase sharply. In the classical limit ($v\ll c$), the kinetic energy per neutron is calculated by $M_{n}^{\star}v^2/2$. Thus the kinetic energy per unit volume is written as,
\begin{equation}\label{equ:kinetic_energy}
\begin{aligned}
  E_{\rm{kin}}=\frac{\hbar^2\rho_n}{8M_{n}^{\star}r^2},
\end{aligned}
\end{equation}
where $\rho_n$ is the local density of neutron within the WS cell.
When the kinetic energy exceeds the condensation energy of superfluid neutrons, the neutron Cooper pairs are destroyed, causing the transition of superfluid neutrons to normal neutrons. Consequently, the vortex has a core that contains normal neutrons. The condensation energy per unit volume is
\begin{equation}\label{equ:condensation_energy}
\begin{aligned}
  E_{\rm{cond}}=-\frac{3\Delta_n^2\rho_n}{8e_{n}},
\end{aligned}
\end{equation}
where $e_n=\sqrt{k_{\mathrm{F},n}^2+M_{n}^{\star2}}-M_{n}^{\star}$ is the Fermi energy which subtract the (in-medium) rest mass and is equivalent to the role of Fermi energy in the non-relativistic case.

The balance of Eq. (\ref{equ:kinetic_energy}) and Eq. (\ref{equ:condensation_energy}) gives the transition radius $R_t=R_t(z)$, which is a function of the vortex axis $z$. According to the above, the local energy cost $E_{\rm{cost}}$ within a vortex is:
\begin{equation}\label{equ:Ecost}
E_{\rm{cost}}=\left\{
\begin{aligned}
  &\frac{\hbar^2\rho_n}{8M_{n}^*r^2},~r>R_t(z) \\
  &\frac{3\Delta_n^2\rho_n}{8e_{n}},~r\leq R_t(z),
\end{aligned}
\right.
\end{equation}
A widely used quantity to represent the vortex core size is the coherence length $\xi$,
\begin{equation}\label{equ:coherencelength}
\begin{aligned}
  \xi = \frac{k_{\rm{F,\infty}}}{\pi M_{\infty}^*\Delta_{\infty}}.
\end{aligned}
\end{equation}
The transition radius, $R_{\infty}$, in the neutron sea is slightly larger than the coherence length $\xi$. The neutron velocity $v$ is always much less than $c$ for $r>R_\infty$ or $r>\xi$, the application of the classical limit in Eq. (\ref{equ:kinetic_energy}) is reasonable.

In Fig. \ref{fig:size}, we also present the neutron vortex core size, i.e., the coherence length $\xi$, as a function of $\rho_{\mathrm{B}}$. Since $\xi$ is inversely proportional to $\Delta_{\infty}$, the coherence length decreases at lower density and then increases at higher density. Stronger polarization results in the smaller $\Delta_{\infty}$ and the larger $\xi$, which in turn gives the longer capture radius in estimates of the pinning force; see \cite{Seveso2016_MNRAS455-3952}. The dependence of these sizes on $\rho_{\mathrm{B}}$, $L_0$, and polarization $\beta$ significantly affect the pinning energy and pinning force.

Now we calculated the total energy cost in the IP and NP configurations. At a given density, we set up a cylindrical container with a radius of $R_{\rm{ws}}$ and a height of $2R_{\rm{ws}}$. The IP configuration is obtained after this container is filled with free neutrons and then a vortex is created with its axis coinciding with the axis of the cylinder; if we place the spherical approximated WS cell at the center of this container, while filling the remaining space with free neutrons and then creating a vortex in the same position as the IP configuration, we get the NP configuration. We label $\rho_n^{\rm{IP}}(\rho_{\mathrm{B}};\rho,\theta,z)$ and $\rho_n^{\rm{NP}}(\rho_{\mathrm{B}};\rho,\theta,z)$ as the neutron density distributions in the IP and NP configurations. Here we need to transfer the 1D density distribution within the WS cell to the density distribution in cylindrical coordinate. We perform the same operations for $\Delta_n$, $M_n^{\star}$ and $e_n$ as for $\rho_n$. The pinning energy is obtained by integrating the local difference between the local energy cost $E_{\rm{cost}}^{\rm{IP}}$ and $E_{\rm{cost}}^{\rm{NP}}$,
\begin{equation}\label{equ:Ep}
\begin{aligned}
  E_{\rm{p}}(\rho_{\mathrm{B}})=2\pi\int_{0}^{2R_{\rm{ws}}}\int_{0}^{R_{\rm{ws}}}\left(
                     E_{\rm{cost}}^{\rm{IP}}-E_{\rm{cost}}^{\rm{NP}}\right)\rho\mathrm{d}\rho\mathrm{d}z.
\end{aligned}
\end{equation}
where $2\pi$ is derived from the integration of $\theta$ due to the axial symmetry.

Note that our calculations only consider the case of vortex pinning to a single site, without accounting for the effects of the actual nuclear lattice in the inner crust. As shown in Fig. \ref{fig:cubic}, when the density approaches the crust-core transition density, the distance between two adjacent droplets becomes less than the size the vortex core. The calculation of energy cost for the IP configuration loses the energy contributions from the neighboring droplets. A more detailed calculation of the pinning energy within the nuclear lattice will be addressed in our future work.

The pinning force per unit length $f_{\rm{pin}}$ can be estimated \citep{Seveso2016_MNRAS455-3952} by using the pinning energies and properties of WS cells that we have calculated.
This estimation takes into account displacement of the nuclei and averages over all possible orientations of polycrystalline structure of the lattice with respect to the vortex. An extra parameter presented in the calculation is the vortex length $l$, we choose $l=5000R_{\rm{ws}}$ in this work, and a brief discussion of the effects of different $l$ on the calculations can be found in Sec. 3.3.

\begin{figure}
\centering
\includegraphics[width=0.45\textwidth]{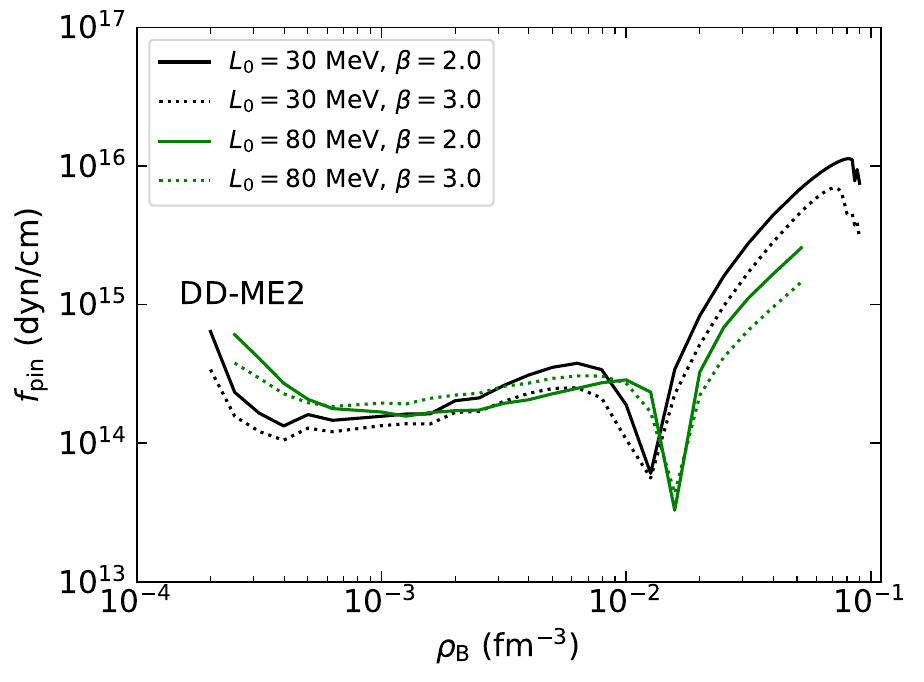}
\caption{Pinning forces calculated by DD-ME2 with $L_0=$ 30, 80 MeV and $\beta=$ 2, 3 from the outer-inner crust transition density $\rho_{\rm{oi}}$ to the crust-core transition density $\rho_{\rm{cc}}$. }
\label{fig:fpin_low}
\end{figure}
\begin{figure*}
\centering
\includegraphics[width=0.9\textwidth]{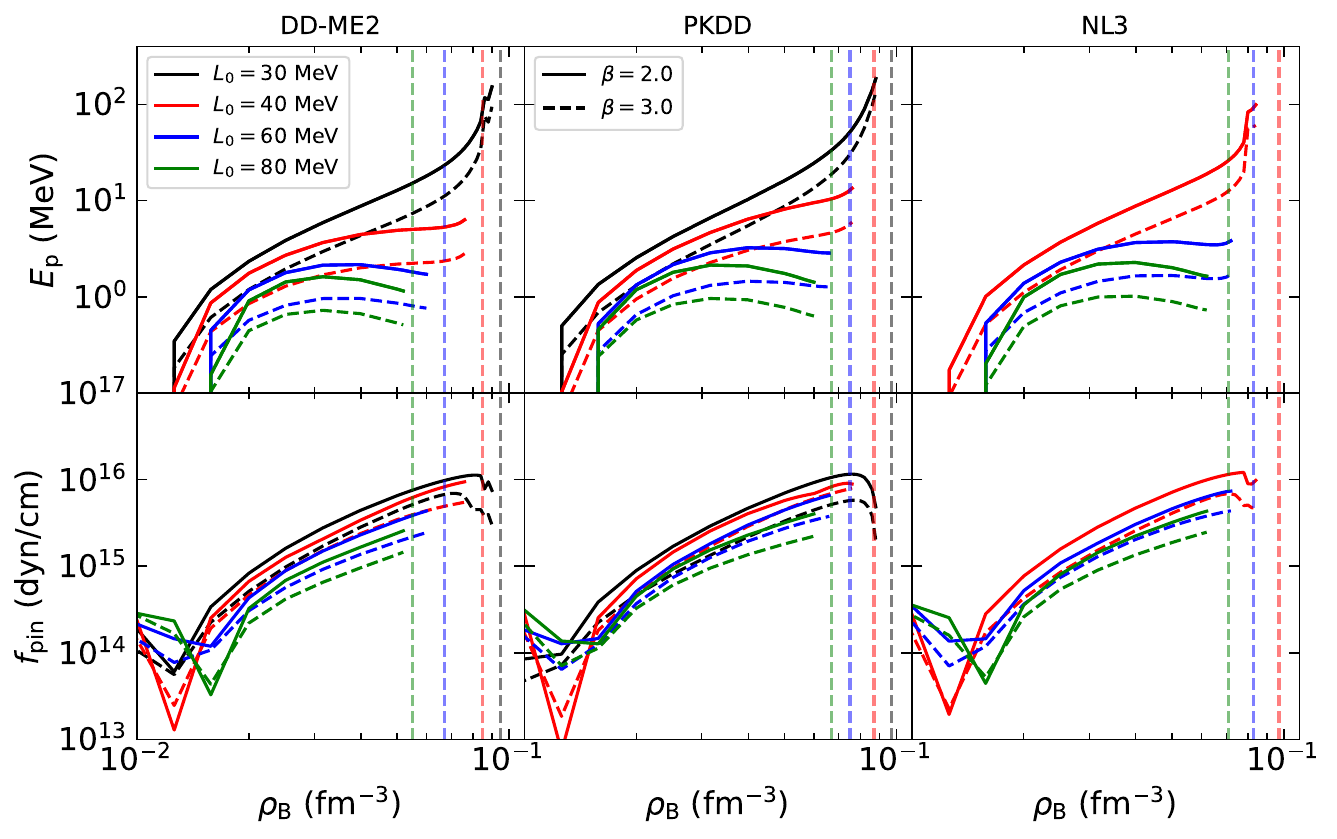}
\caption{Pinning energies (upper panels) and pinning forces (lower panels) as a function of $\rho_{\mathrm{B}}$ for DD-ME2 (left panels), PKDD(middle panels), and NL3 (right panels) with $L_0=$30, 40, 60, 80 MeV and $\beta=$ 2.0 and 3.0. The core-crust transition densities are depicted by the corresponding dashed lines. We only present the results where $E_{\rm{p}}>$ 0, which correspond to the configurations of vortex pinning on the droplet. The results for $E_{\rm{p}}<$ 0 indicate that the pinning force is much weaker than the pinning force just close to the core-crust transition density.}
\label{fig:Ep}
\end{figure*}

\subsection{The snowplow model for pulsar glitch}\label{sec:theor_snowplow}
For completeness, we briefly introduce the snowplow model for pulsar glitch in this section.
According to the snowplow scenario \citep{Pizzochero2011_ApJ743-L20,Seveso2012_MNRAS427-1089}, neutron vortices are assumed to be globally rigid and parallel to the rotation axis of the pulsar, with the ends of the vortices terminating at the interface between the outer and inner crust. The balance between the pinning force $F_{\rm{pin}}$,
\begin{equation}\label{equ:Fpin}
\begin{aligned}
  F_{\mathrm{pin}}(r)=2\int_{0}^{l(r)/2}f_{\mathrm{pin}}(\sqrt{r^{2}+z^2})\mathrm{d}z,
\end{aligned}
\end{equation}
and the Magnus force $F_{\rm{mag}}$,
\begin{equation}\label{equ:Fmag}
\begin{aligned}
  F_{\mathrm{mag}}(r)&=F_{\mathrm{mag}}^*\Delta\Omega \\
  &=2\kappa r\Delta\Omega\int_{0}^{l(r)/2}\rho_{\mathrm{s}}(\sqrt{r^2+z^2})\mathrm{d}z,
\end{aligned}
\end{equation}
acting the whole vortex line results in the critical angular velocity lag $\Delta\Omega_{\rm{cr}}=F_{\rm{pin}}/F_{\rm{mag}}^*$, which is required for vortex unpinning, between the superfluid component and the normal component. Here, $r$ and $z$ are the distances from the rotation axis and the equatorial plane, respectively. $l(r)=2\sqrt{R_{\rm{oi}}^2-r^2}$ is the length of vortex line with the inner crust radius $R_{\rm{oi}}$. The
constant $\kappa$ is the quantum of circulation of a neutron fluid. $\rho_{\rm{s}}$ is the local neutron superfluid mass density. We assume that the neutrons are superfluid throughout the star, e.g., $^1S_0$ pairing in the inner crust and $^3P_2$ pairing in the core. Hence $\rho_{\rm{s}}$ is written by $\rho_{\rm{s}}=(1-Y_p)\rho$ with the proton fraction $Y_p$ and the mass density $\rho$.

Due to the dependence of $\Delta\Omega_{\rm{cr}}$ on $r$, the continuously unpinning and repinning of vortices result in a collective outward radial motion of the vortices. Vortices are gradually accumulated in regions of higher $\Delta\Omega_{\rm{cr}}$, ultimately concentrating in the thin sheet near $\Delta\Omega_{\rm{cr}}^{\rm{max}}=\Delta\Omega_{\rm{cr}}(r_{\mathrm{max}})$, where $\Delta\Omega_{\rm{cr}}$ reaches its maximum value at $r=r_{\mathrm{max}}$. When $\Delta\Omega_{\rm{cr}}^{\rm{max}}$ is reached, a large number of vortices depin and transfer the angular momentum to the normal component in the region where $r_{\mathrm{max}}<r<r_{\mathrm{o}}$, causing a vortex avalanche that triggers the occurrence of a glitch. $r_{\mathrm{o}}$ and $r_{\mathrm{max}}$ are the inner and outer boundaries of the angular momentum transfer region, respectively.

For an axially symmetric system, the angular velocity $\Omega_{\rm{s}}$ of the superfluid component is proportional to the number $N(r)$ of vortices enclosed in a cylindrical region with radial distance $r$. Hence the number of vortices accumulated in peak of $\Delta\Omega_{\rm{cr}}^{\rm{max}}$ is:
\begin{equation}\label{equ:Npeak}
\begin{aligned}
  N_{\rm{v}}=\frac{2\pi}{\kappa}r_{\mathrm{max}}^2\Delta\Omega_{\rm{cr}}^{\rm{max}}.
\end{aligned}
\end{equation}
Only $N_{\rm{v}}$ vortices respond to the glitch, and the vortices in the region of $r_{\mathrm{max}}<r<r_{\mathrm{o}}$ are assumed to be completely removed by the Magnus force just before a glitch. Then we can calculate the angular momentum transfer by applying the $\mathrm{d} L=\Omega_{\rm{s}} \mathrm{d}  I_{\rm{s}}$,
\begin{equation}\label{equ:dL}
\begin{aligned}
  \mathrm{d} L = 2\kappa N_{\mathrm{v}}\int_{r_{\mathrm{max}}}^{r_{\mathrm{o}}}r\mathrm{d} r\int_{0}^{\sqrt{r_{\mathrm{o}}^2-r^2}}\rho_{\mathrm{s}}(\sqrt{r^2+z^2})\mathrm{d} z.
\end{aligned}
\end{equation}

To evaluate the jump in angular velocity of normal component $\Delta\Omega_{\rm{gl}}$, the fraction $Y_{\rm{gl}}$ of core superfluid coupled to the normal component is introduced, but it cannot be inferred by the snowplow model. The angular momentum conservation gives:
\begin{equation}\label{equ:DeltaOmegaJump}
\begin{aligned}
  \Delta\Omega_{\rm{gl}} = \frac{\Delta L}{I_{\rm{tot}}[1-Q(1-Y_{\rm{gl}})]},
\end{aligned}
\end{equation}
where $Q$ is the ratio of the total moment of inertia of superfluid to the total moment of inertia of the star. The relative deceleration of the normal component immediately after the glitch reads:
\begin{equation}\label{equ:Step}
\begin{aligned}
  \frac{\Delta\dot{\Omega}_{\rm{p}}}{\dot{\Omega}_{\rm{p}}}=\frac{Q(1-Y_{\rm{gl}})}{1-Q(1-Y_{\rm{gl}})}.
\end{aligned}
\end{equation}

Vela 2000 glitch gives $\Delta\Omega_{\rm{gl}}=2.2\times10^{-4}$ rad/s and $\Delta\dot{\Omega}_{\rm{p}}/\dot{\Omega}_{\rm{p}}=18\pm6$. We evaluate $Y_{\rm{gl}}$ by fitting Eq. (\ref{equ:DeltaOmegaJump}) to observation, then calculate $\Delta\dot{\Omega}_{\rm{p}}/\dot{\Omega}_{\rm{p}}$ by Eq. (\ref{equ:Step}) and compare to the observation. The effective interactions that yield results inconsistent with observations can be excluded.

\section{Results and discussion}\label{sec:res}

\subsection{Dependences of pinning energy and pinning force on nuclear parameters}

Using the compositions of the inner crust obtained from self-consistent calculations, we calculate the pinning energies $E_{\rm{p}}$ within the inner crust under different $L_0$ and polarization strengths $\beta$ using Eq. (\ref{equ:Ep}).

Fig. \ref{fig:fpin_low} shows the pinning force per unit length calculated by DD-ME2 with $L_0=$ 30 and 80 MeV throughout the inner crust. We can see that the lower pinning energy of the IP configuration results in a weak pinning region at the top of the inner crust, while the NP configuration creates a strong pinning region near the crust-core transition density. The pinning force per unit length in the weak pinning region is about an order of magnitude smaller than that in the strong pinning region.
Based on our calculations, $E_{\rm{p}}$ versus $\rho_{\mathrm{B}}$ exhibit a similar trend for different effective interactions and polarization strengths, with the specific values depending on these factors.

The upper panels in Fig. \ref{fig:Ep} present the pinning energy calculated by DD-ME2, PKDD, and NL3 with $L_0=$ 30, 40, 60, and 80 MeV over a density range of 0.01--0.10 fm$^{-3}$. For all effective interactions, the pinning undergoes a transition from IP configuration to NP configuration in the range of 0.01--0.02 fm$^{-3}$. Roughly speaking, independent of the effective interaction, in the density range of $\rho_{\rm{oi}}$ to 0.01 fm$^{-3}$, vortices prefer to pin in the neutron sea between two neighboring droplets; however, in the density range from 0.02 fm$^{-3}$ to $\rho_{\rm{cc}}$, vortices are more likely to pin on the droplets. For all effective interactions and polarization strengths, the pinning energy for the IP configuration typically ranges between $-2$ and 0 MeV, while the pinning energy for the NP configuration can reach several tens of MeV to hundreds of MeV, as we can see in Fig. \ref{fig:Ep}.

Note that the transition from IP configuration to NP configuration is also observed in previous semi-classical calculation \citep{Pizzochero1997_PRL79-3347}, although the pinning strengths in our work are weaker than those in \cite{Pizzochero1997_PRL79-3347} due to the inclusion of polarization. Quantum calculations suggest that NP configuration is more favorable at the top of the inner crust, while the pinning configuration may depend on the effective interaction and polarization at the bottom of the inner crust \citep{Avogadro2007_PRC75-012805,Klausner2023_PRC108-035808}. The difference between two approaches is that quantum calculation takes into account the shell effects and the much larger vortex radius in NP configuration compared to the semiclassical approach, see discussions also in \cite{Avogadro2008_NPA811-378}.

\begin{figure}[t]
\centering
\includegraphics[width=0.45\textwidth]{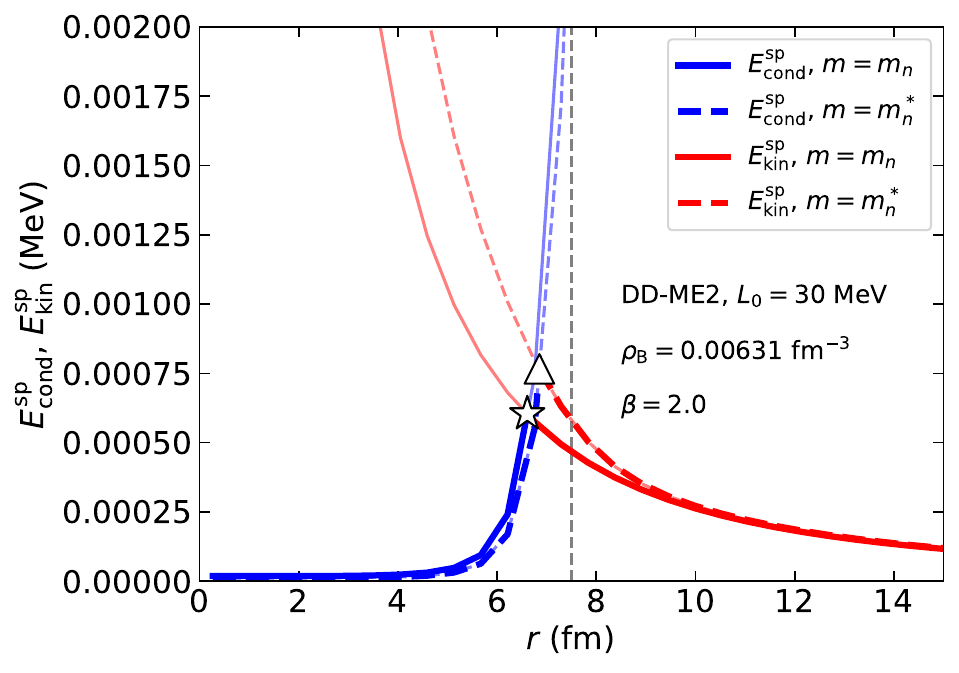}
\caption{The condensation energy per particle $E_{\rm{cond}}^{\rm{sp}}$ and kinetic energy per particle $E_{\rm{kin}}^{\rm{sp}}$ as a function of $r$ in the cases of bare neutron mass and Dirac neutron mass. The effective interaction is DD-ME2 with $L_0=30$ MeV. The matter distribution is calculated at $\rho_{\mathrm{B}}=$0.00631 fm$^{-3}$. The polarization is taken as $\beta=2.0$. }
\label{fig:mstar}
\end{figure}

Next, we discuss the dependence of the pinning energy and pinning force per unit length on the effective interaction. In the upper panels of Fig. \ref{fig:Ep}, we observe that the pinning energy shows a slight dependence on the interactions in the isoscalar channel, but a significant dependence on the interaction in the isovector channel. When the polarization strength is unchanged, a higher $L_0$ corresponds to a lower pinning energy, and a clear maximum in the pinning energy can be observed before the crust-core transition density. As $L_0$ decreases, the pinning energy increases, the maximum of the pinning energy shifts to higher densities, and disappears when $L_0$ is below 40 MeV. The pinning energy corresponding to a smaller $L_0$ may exceed 100 MeV at the bottom of the inner crust, such extreme large pinning energy is not found in previous works and can be explained by the size of the inhomogeneous structure: On the one hand, due to the strong surface tension at small $L_0$, the droplet at the bottom of the inner crust can maintain a defined inhomogeneous structure at large $R_{\rm{d}}$ and $R_{\rm{ws}}$; On the other hand, the large size leads to a large-scale volume integral in Eq. (\ref{equ:Ep}), so that the small local energy cost differences between IP and NP configurations can accumulate into a large pinning energy. The dependence of the pinning force per unit length on the effective interaction is similar to that of the pinning energy. However, due to the introduction of size effects in the Eq. (21) of \cite{Seveso2016_MNRAS455-3952}, the compact structure at the bottom of the inner crust leads to a reduction in the pinning force per unit length. Therefore, in contrast to the pinning energy, the maximum of the pinning force per unit length can be observed at small $L_0$ and disappears at high $L_0$. We also find that the dependence of pinning force per unit length on $L_0$ is affected by polarization.

The dependence of the pinning energy and pinning force per unit length on polarization strength is direct, i.e., stronger polarization leads to a reduction in both the pinning energy and pinning force per unit length. Since the condensation energy is proportional to $\Delta^2$, strong polarization results in a smaller condensation energy. The balance between the local kinetic energy and the local condensation energy under strong polarization is achieved at a larger $r$, and therefore a larger vortex core is obtained. In the case of strong polarization, the vortex core in the NP configuration is able to enclose the entire droplet. As a result, the difference in the kinetic energy contributions between the IP and NP configurations tends to disappear, and then the pinning energy is lowered. Polarization introduces an additional size parameter, i.e., coherence length $\xi$, in the estimation of the pinning force per unit length; the dependence of the pinning force on $L_0$ under different polarization may be non-monotonic. For example, the pinning force per unit length calculated with PKDD with $L_0=40$ MeV under strong polarization is significantly larger than that calculated with PKDD with $L_0=30$ MeV, which is in contrast to the case of weak polarization.

It is important to note that, as shown in Fig. \ref{fig:Ep}, we do not present the pinning energy and pinning force per unit length at several densities close to the crust-core transition. The reason is that our calculations fail to yield a stable droplet structure in these densities, which strongly suggests that the non-spherical structures should be included in future work.

It is worth mentioning that, treating vortex avalanches as a state-dependent Poisson process, \cite{Melatos2023_ApJ948-106} performed a Bayesian analysis of the pulsar glitch rates to obtain the maximum $\Delta\Omega_{\rm{cr}}$ and $f_{\rm{pin}}$. Their estimate of $f_{\rm{pin}}$ is independent of our calculations based on nuclear theory. Our calculations indicate a pinning force that is at least three times smaller than the value obtained by \cite{Melatos2023_ApJ948-106}.
Additionally, it should be noted that \cite{Melatos2023_ApJ948-106} did not consider other state-dependent Poisson processes that do not involve vortex avalanches and nuclear pinning, such as starquake, nor did they account for the vortex pinning in the core and general relativistic effects.

In Eqs. (\ref{equ:kinetic_energy}) and (\ref{equ:condensation_energy}), we have replaced the bare mass and Fermi energy used in the non-relativistic calculations with the Dirac effective mass and the exceeded Fermi energy, respectively, thus the nuclear medium effects are considered in our work. In Fig. \ref{fig:mstar}, we show the single-particle kinetic energy $E_{\rm{kin}}^{\rm{sp}}$ and condensation energy $E_{\rm{cond}}^{\rm{sp}}$ distributions within the WS cell in the cases of with and without nuclear medium effects. The effective interaction, polarization, and density are taken as DD-ME2 with $L_0=30$ MeV, $\beta=2.0$, and 0.00631 fm$^{-3}$, respectively. We have assumed that the density and pairing gap distributions within the WS cell remain unchanged in both cases. We find that, because of the reduction in nucleon mass, the nuclear medium effects lead to a larger kinetic energy contribution, a smaller condensation energy contribution, and a larger vortex core size. Furthermore, we also calculate the pinning energy at $\rho_{\mathrm{B}}=0.00631$ fm$^{-3}$ (IP configuration) and $\rho_{\mathrm{B}}=0.062$ fm$^{-3}$ (NP configuration). For the case without nuclear medium effects, the pinning energies are $-1.20$ MeV and 22.97 MeV, respectively; After the nuclear medium effects are considered, the pinning energies change to $-1.05$ MeV and 19.34 MeV. Thus, the smaller pinning energy is expected for both IP and NP configurations when the nuclear medium effects are considered.

\begin{figure}[t]
\centering
\includegraphics[width=0.45\textwidth]{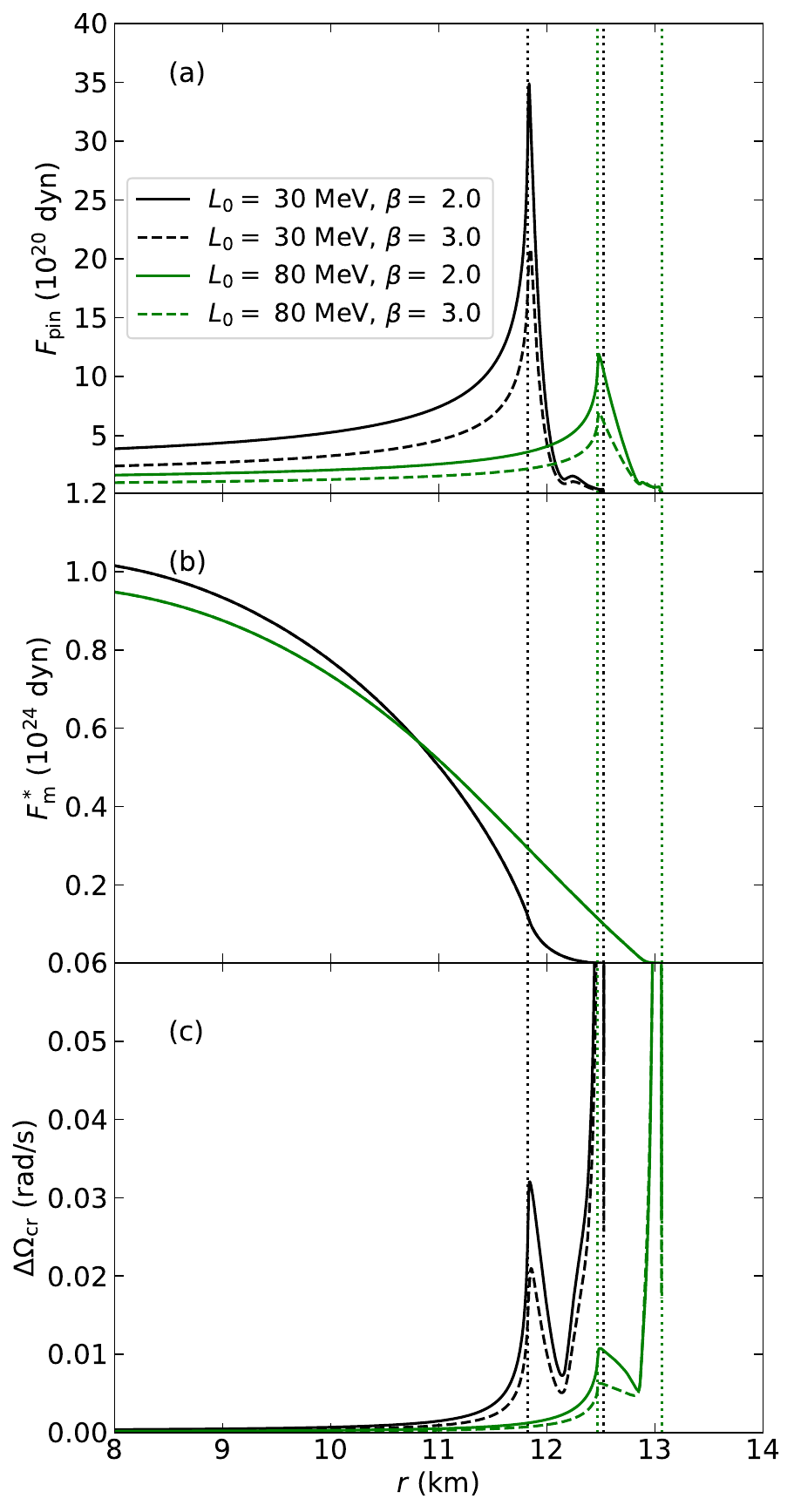}
\caption{(a) Pinning force $F_{\rm{pin}}$, (b) Magnus force $F_{\rm{mag}}^*$, and (c) Critical angular momentum lag $\Delta\Omega_{\rm{cr}}$  as the function of the distance of the vortex line to the rotational axis $r$. The black and green lines represent the results for $L_0=$ 30 and 80 MeV, while solid and dashed lines corresponding to the polarization strengths $\beta=$ 2.0 and 3.0, respectively. The effective interaction in isoscalar channel is DD-ME2. We calculate the structure of 1.4 $M_{\odot}$ NS with $R=13.01$ km for $L_0=$ 30 MeV and $R=13.60$ km for $L_0=$ 80 MeV. Since we mainly focus on the pinning properties of the inner crust, the curves for $r<8$ km are not shown, and they are similar to Fig. 2 and Fig. 3 in \citet{Shang2021_ApJ923-108}. Regarding $\Delta\Omega_{\rm{cr}}$, we observed a extreme large lag near the outer boundary of the inner crust. Due to the magnitude of this lag being nearly unattainable, we truncated $\Delta\Omega_{\rm{cr}}$ to below 0.06 rad/s in the panel (c). }
\label{fig:snowplow}
\end{figure}

\subsection{Constraints from the 2000 Vela glitch}

\begin{figure*}[t]
\centering
\includegraphics[width=0.9\textwidth]{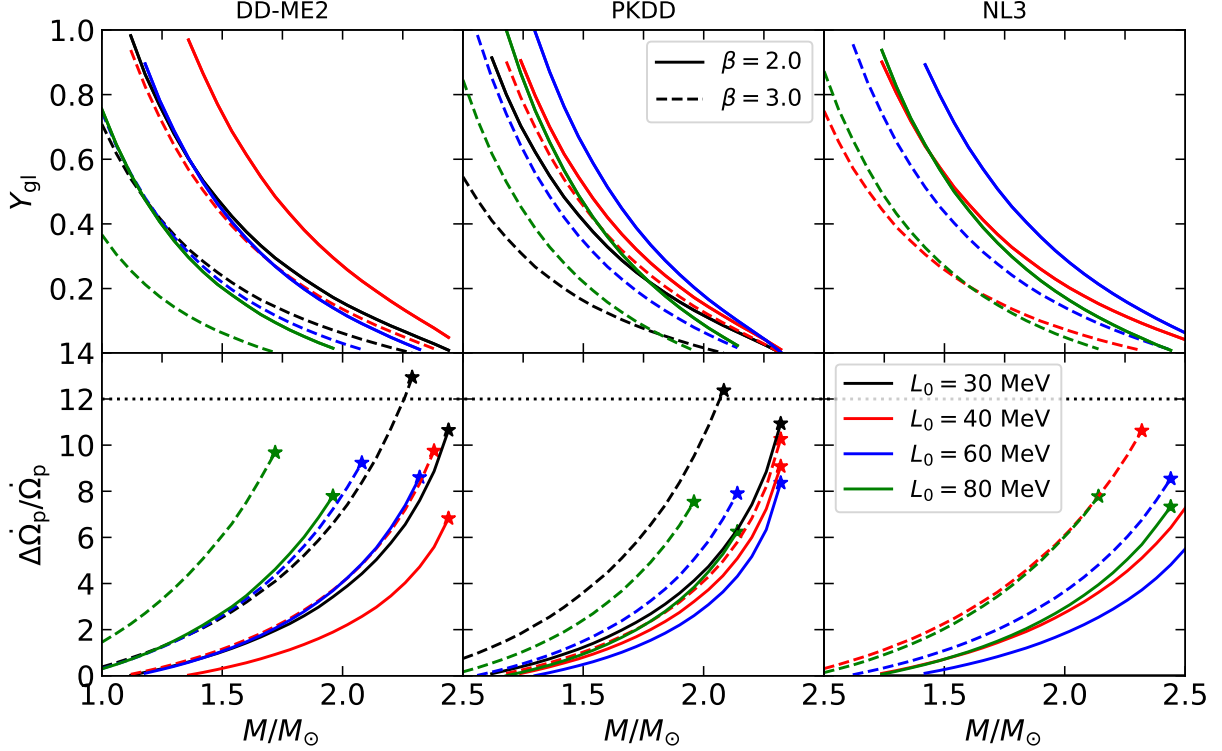}
\caption{Fraction of the coupled vorticity $Y_{\rm{gl}}$ (upper panels) and observed step in spin-down rate $\Delta\dot{\Omega}_{\rm{p}}/\dot{\Omega}_{\rm{p}}$ (lower panels) as a function of NS mass $M/M_{\odot}$ for DD-ME2 (left panels), PKDD(middle panels), and NL3 (right panels) with $L_0=$30, 40, 60, 80 MeV and $\beta=$ 2.0 and 3.0. The dotted line represent the upper bound of measured value for the 2000 Vela glitch, together with the $1\sigma$ deviation.}
\label{fig:vela}
\end{figure*}

\begin{figure}[t]
\centering
\includegraphics[width=0.45\textwidth]{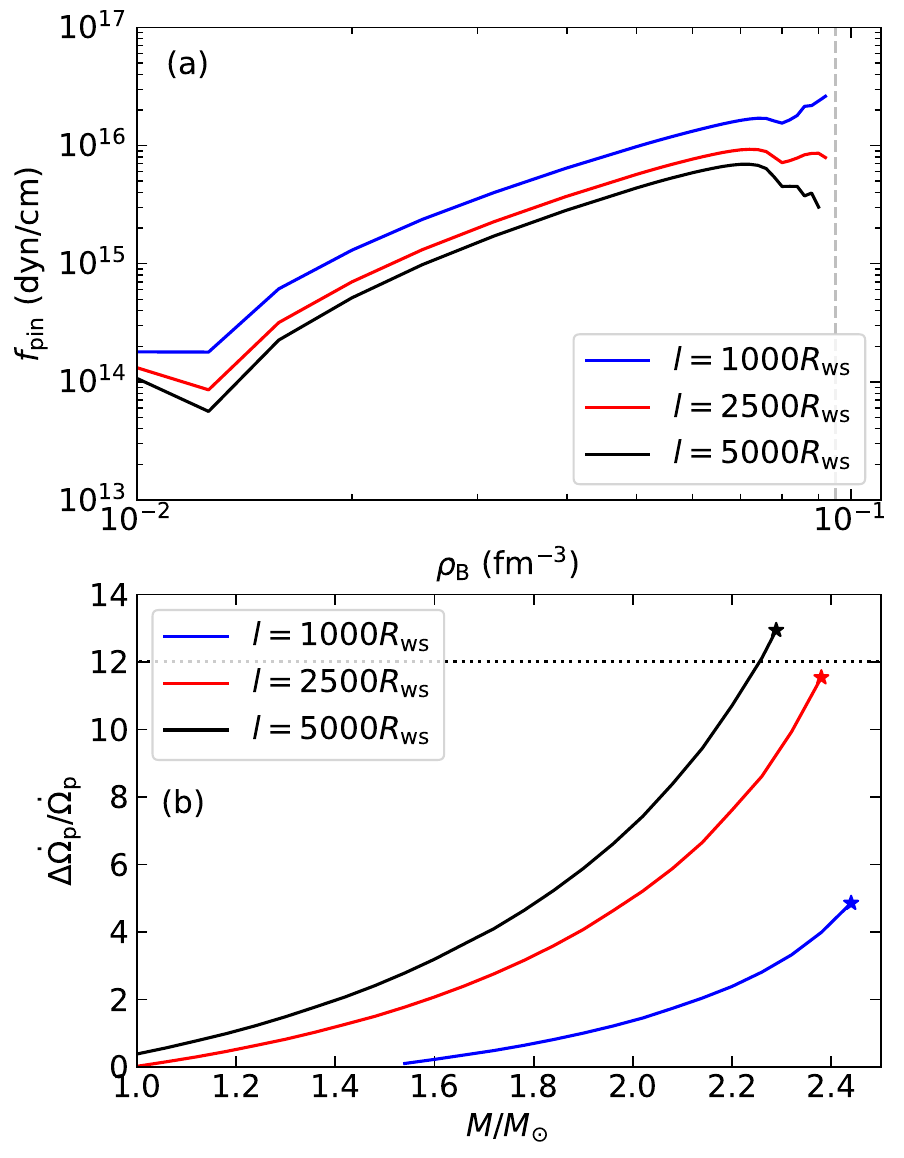}
\caption{(a)Pinning forces $f_{\rm{pin}}$ and (b) observed step in spin-down rate $\Delta\dot{\Omega}_{\rm{p}}/\dot{\Omega}_{\rm{p}}$ for different vortex length. The effective interaction is DD-ME2 with $L_0=30$ MeV. The polarization is taken as $\beta=3.0$.}
\label{fig:tension}
\end{figure}

\begin{figure}[t]
\centering
\includegraphics[width=0.45\textwidth]{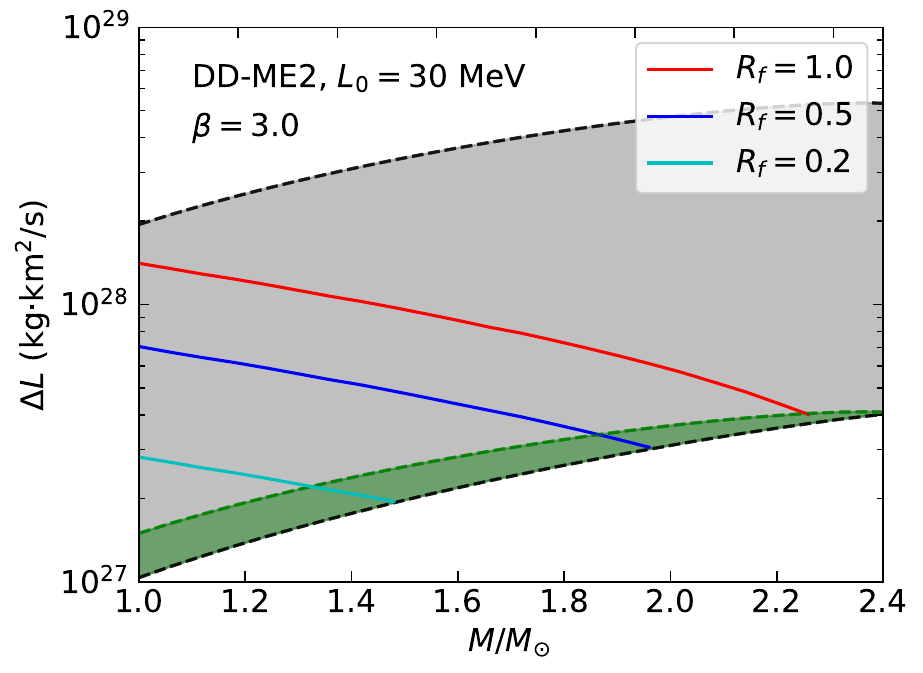}
\caption{The angular momentum transfer $\Delta L$ as a function of NS mass $M$, for three values of the scaling factor $R_f$. The snowplow model requires $Y_{\rm{gl}}$ to lie between 0 and 1. The angular momentum transfer that satisfies the model requirement is denoted by gray shadow, with two black dashed lines representing the upper and lower limits of the angular momentum transfer. The green dashed line shows the upper limit of the angular momentum transfer that is consistent with the observations of Vela 2000, and the angular momentum transfer that simultaneously satisfies both the model requirement and observations is represented by green shadow.
}
\label{fig:DL}
\end{figure}

By solving the TOV equations, we can obtain the interior density profile of NSs. The pinning force per unit length as a function of $\rho_{\mathrm{B}}$ can be mapped to the pinning force profile within the NSs. Furthermore, the pinning energy acting on an entire vortex $F_{\rm{pin}}$ can be derived from Eq. (\ref{equ:Fpin}). In Fig. \ref{fig:snowplow}, we show $F_{\rm{pin}}$, the Magnus force $F_{\rm{mag}}^*$, and critical angular velocity lag $\Delta\Omega_{\rm{cr}}$ at various distances $r$ from the rotation axis. The maximum of $f_{\rm{pin}}$ is located near the crust-core transition density, thus the peak of $F_{\rm{pin}}$ is also very close to the interface between the crust and core; $F_{\rm{mag}}^*$ is relatively smooth near the crust-core interface, leading to a peak in $\Delta\Omega_{\rm{cr}}$ near this interface. In the Fig. \ref{fig:snowplow}, although the shapes of $F_{\rm{pin}}$, $F_{\rm{mag}}^*$, and $\Delta\Omega_{\rm{cr}}$ have no significant differences in different cases, while their magnitudes and positions of extrema depend on $L_0$ and polarization $\beta$. From the upper panel of Fig. \ref{fig:snowplow}, as $L_0$ increases, the NS radius increases and the pinning force per unit length decreases, resulting in that $F_{\rm{pin}}$ decreases and the peak of $F_{\rm{pin}}$ shifts farther from the rotation axis. Polarization reduces the pinning energy, thereby further decreasing $F_{\rm{pin}}$, but the position of the peak remains unchanged. From the middle panel of Fig. \ref{fig:snowplow}, a larger $L_0$ corresponds to a larger NS radius, a smaller fraction of superfluid neutrons in the core, and a larger fraction of superfluid neutrons in the crust. Therefore, in the case of $L_0$ is larger, vortices closer to the rotation axis experience weaker $F_{\rm{mag}}^*$; while those in the inner crust, i.e., vortices farther from the axis, experience the stronger $F_{\rm{mag}}^*$ . Since polarization has no effect on the bulk EoSs, it does not affect the Magnus force. The balance between $F_{\rm{pin}}$ and $F_{\rm{mag}}^*$ gives the lower panel of Fig. \ref{fig:snowplow}. It is clear that the dependences of $\Delta\Omega_{\rm{cr}}$ on $L_0$ and polarization is consistent with those of $F_{\rm{pin}}$.

However, we can find two peaks in $\Delta\Omega_{\rm{cr}}$ profiles for all cases. As we can see in Fig. \ref{fig:fpin_low}, the pinning configuration is changed from IP to NP at about 0.01 fm$^{-3}$, a deep valley in the pinning force per unit length divides the crust into weak and strong pinning regions. The strong pinning region gives rise to the peak in $\Delta\Omega_{\rm{cr}}$ at $r=R_{\rm{L}}^*$ near the crust-core interface, while the weak pinning region results in a peak at $r=R_{\rm{R}}^*$ closer to the outer boundary of the inner crust. Although $F_{\rm{pin}}$ is smaller near the outer boundary of the inner crust, $F_{\rm{mag}}^*$ experienced by vortices in this region is even smaller, creating the peak in $\Delta\Omega_{\rm{cr}}$ whose magnitude is much larger than that of peak derived from the strong pinning region.

We assume that the angular velocity lag $\Delta\Omega_{\rm{cr}}$ between the normal component and the superfluid component inside the NSs changes smoothly. A large number of vortices accumulate in the shell near the peak close to the crust-core interface. When these vortices are suddenly unpinned due to $\Delta\Omega_{\rm{cr}}$ exceeding its maximum value, a glitch occurs and $\Delta\Omega_{\rm{cr}}$ decreases. The mechanism means that the vortices near the peak at the outer boundary of the inner crust cannot be unpinned due to their extreme large lag. Therefore, the weak pinning region of an NS have a negligible contribution to the triggering of glitch and its successive relaxation. We further assume that vortices in the region where $R_{\rm{L}}^*<r<R_{\rm{M}}^*$ have been completely removed by the Magnus force after the glitch is triggered, where $R_{\rm{M}}^*$ is the location of the minimum of the $\Delta\Omega_{\rm{cr}}$ profile (between two peaks); then the angular momentum transfer can be immediately evaluated via Eq. (\ref{equ:dL}) and by setting $r_{\mathrm{max}}=R_{\rm{L}}^*$ and $r_{\mathrm{o}}=R_{\rm{M}}^*$.

By fitting the jump in angular velocity of normal
component $\Delta\Omega_{\rm{gl}}$ of Vela 2000, $Y_{\rm{gl}}$ and $\Delta\dot{\Omega}_{\rm{p}}/\dot{\Omega}_{\rm{p}}$ can be obtained through Eqs. (\ref{equ:DeltaOmegaJump}) and (\ref{equ:Step}), respectively. Fig. \ref{fig:vela} illustrates $Y_{\rm{gl}}$ and $\Delta\dot{\Omega}_{\rm{p}}/\dot{\Omega}_{\rm{p}}$ as a function of NS mass $M$ for different effective interactions and polarizations. The end of the $\Delta\dot{\Omega}_{\rm{p}}/\dot{\Omega}_{\rm{p}}$ versus $M$ is determined by $Y_{\rm{gl}}$, which corresponds to the fraction of core superfluid coupled to the crust being zero. From the lower panels of Fig. \ref{fig:vela}, the lower limit of observation can be satisfied for DD-ME and PKDD with the same symmetry energy slope $L_0=$ 30 MeV and strong polarization $\beta=3.0$. The Vela mass constrained by the observation is 2.255–-2.290 $M_{\odot}$ for DD-ME2 and 2.069–-2.084 $M_{\odot}$ for PKDD, respectively. For NL3, none of EoSs satisfy the observations under weak or strong polarization. However, the interaction with $L_0=$ 40 MeV under strong polarization is very close to the lower limit. Therefore, we expect that an interaction with $L_0<$ 40 MeV, which its EoS is also monotonic, can satisfy the observation. From the upper panels of Fig. \ref{fig:vela}, the $Y_{\rm{gl}}$ of NSs matching the observations are below 1$\%$, this means that the coupling between the core superfluid neutrons and the crust is very weak. In conclusion, the 2000 Vela glitch constrains $L_0$ to be below 40 MeV and polarization to be strong $\beta\sim$ 3.0, which are consistent with previous works \citep{Hooker2013_JPCS420-012153,Shang2021_ApJ923-108}. Interestingly, our results suggest that Vela is a massive NS, as discussed in greater detail in Sec.~\ref{sec:strength}.

\cite{Haskell2013_ApJ764-L25} also applied the snowplow model to the 2000 Vela glitch
assuming strong pinning between the flux tubes and neutron vortices.
Our results can be compared with theirs using the fraction of pinned vorticity in the core set to zero.
In \cite{Haskell2013_ApJ764-L25}, the results calculated with most EoSs and $Y_p$ match the observed value. However, our results are hard to reconcile with the observed data, and fitting them is
only possible for several effective interactions with $L_0<40$ MeV and strong polarization in a narrow interval of masses. One reason is that the location of the maximum $\Delta\Omega_{\rm{cr}}$ is close to the bottom of the inner crust, as shown in Fig. \ref{fig:snowplow}.
From Eq. (\ref{equ:dL}), the angular momentum transfer increases when the crust region supported for the angular momentum transfer is expanded. Therefore, compared to \cite{Haskell2013_ApJ764-L25}, our calculations indicate that the inner crust contains a larger angular momentum transfer region, leading to more angular momentum transfer, finally resulting in an increase in $Y_{\rm{gl}}$ and a decrease in $\Delta\dot{\Omega}_{\rm{p}}/\dot{\Omega}_{\rm{p}}$.

\subsection{The effect of vortex length}

The rigidity length $l$ of vortex is related to its tension. A stronger tension reduces the ability of a vortex to bend and pin to a pinning configuration, resulting in a lower the pinning force.
In previous calculations, we have assumed a vortex length of $l=5000R_{\rm{ws}}$. Based on \cite{Seveso2016_MNRAS455-3952}, we also investigate the impact of $l$ on our results. In Fig. \ref{fig:tension}(a), we show pinning forces as a function of density for $l=1000R_{\rm{ws}},~2500R_{\rm{ws}}$, and $5000R_{\rm{ws}}$. We can see that the pinning force increases as $l$ decreases. For smaller $l$, the pinning force does not show a significant decline near the crust-core density. Fig. \ref{fig:tension}(b) illustrates the constraints from the 2000 Vela glitch for different $l$. Clearly, Vela 2000 does not support a short vortex length, which implies that the vortex tension is strong.

\subsection{The effect of pinning force strength}
\label{sec:strength}
Since EoSs and pinning forces are self-consistently calculated in this work, the 2000 Vela glitch provides a strong constraint on the NS mass. We now temporarily break the relation between EoS and pinning force and discuss the constraints from the 2000 Vela glitch under different pinning force strengths. According to Eqs. (\ref{equ:dL}) and (\ref{equ:DeltaOmegaJump}), when EoS and NS mass are given, the angular momentum transfer $\Delta L$ during a glitch depends only on the distribution of the pinning force within the inner crust. We use the scaling factor $R_f$ to parameterize the pinning force, i.e., $f_{\rm{pin}}(R_f)=R_ff_{\rm{pin}}$, so the angular momentum transfer depends only on $R_f$.

In Fig. \ref{fig:DL}, we present the angular momentum transfer $\Delta L$ as a function of NS mass for different $R_f$. The effective interaction and the polarization are taken as DD-ME2 with $L_0=30$ MeV and $\beta=3.0$, respectively. In Fig. \ref{fig:DL}, the upper and lower black dashed lines correspond to the lower and upper limits of angular momentum transfer; the latter are obtained by setting $Y_{\rm{gl}}=0$ and $Y_{\rm{gl}}=1$ in Eq. (\ref{equ:DeltaOmegaJump}), respectively. The angular momentum transfers that validate the snowplow model are shaded by gray. By substituting the upper and lower limits of $\Delta\dot{\Omega}_{\rm{p}}/\dot{\Omega}_{\rm{p}}$ into Eqs. (\ref{equ:DeltaOmegaJump}) and (\ref{equ:Step}), we can obtain the lower and upper limits of angular momentum transfer that are consistent with the observations. The green dashed line in Fig. \ref{fig:DL} draws the upper limit of angular momentum transfer derived from the observed lower limit. Then, the angular momentum transfer during a glitch should lie within the green region shown in Fig. \ref{fig:DL} in order to simultaneously satisfy the snowplow model requirements and the observations from the 2000 Vela glitch. We can see that the angular momentum transfer decreases as the NS mass increases. This is because the heavier NSs correspond to the smaller radius, which can reduce the moment of inertia available for angular momentum transfer after the vortex avalanche. The angular momentum transfer decreases with decreasing pinning force , allowing lighter NSs with thicker crusts to match observations of the 2000 Vela glitch. As shown in Fig. \ref{fig:DL}, when $R_f$ is reduced from 1 to 0.5 and 0.2, the 2000 Vela glitch constrains its mass from about 2.3 $M_{\odot}$ to 1.9 $M_{\odot}$ and 1.4 $M_{\odot}$, respectively.

\citet{Shang2021_ApJ923-108} found that Vela mass is unexpectedly small. Although the snowplow model is also applied in their work, the pinning force is obtained by fitting $\Delta\Omega_{\rm{cr}}$ evaluated from the inter-glitch time of Vela. The maximum of pinning force is about 10$^{15}$ dyn/cm and is much smaller than the pinning force used in our work. As a result, according to Fig. \ref{fig:DL}, the unexpectedly small mass for Vela was obtained then.

\section{Summary and Perspective}\label{sec:summary}
Pulsar glitch study has the potential to not only provide unique constraints on the dense matter EoS, but also shed light on the composition and state of matter at supranuclear densities.
In this work, to maximise the information extracted from glitch observations, we have demonstrated an attempt to explore nuclear force in the context when all required ingredients are properly represented.
In practice, we refit the effective interactions DD-ME2, PKDD, and NL3 to the nuclear saturation properties, and then they are used to perform the calculations of the unified EoSs, composition, and global properties of NSs in a consistent manner. Using the semi-classical method, the evolutions of associated pinning force acting on the vortex are determined with the dependence to the nuclear parameters and superfluid polarization strength. Taking these microscopic and macroscopic inputs to the snowplow model for pulsar glitch, the constraints to nuclear force and Vela mass from the 2000 Vela glitch are conducted.

We find that vortices favor interstitial pinning with weak strength in the top of the inner crust;  while vortices strongly pinning to nuclei in the bottom. In the isoscalar channel, the nuclear forces have slight impact to the pinning properties, at least for these forces we choose. In the isovector channel, the smaller symmetry energy slope leads to the weaker pinning for interstitial pinning case and to the stronger pinning for nuclear pinning case. The stronger polarization reduces the pinning strength for both interstitial and nuclear pinning cases. The observation of the 2000 Vela glitch constrains the symmetry energy slope to below 40 MeV, and 
indicates that it is a pulsar with a large mass.

In future work, we will further incorporate entrainment effects \citep{Chamel2013_PRL110-011101,Chamel2024_arXiv2412.05599} and fit the improved results to the phenomena at the early-stage of the 2016 Vela glitch \citep{Ashton2019_NA36-844,Pizzochero2020_A&A636-A101}, such as the glitch rise time and the rapid recovery of overshooting. This aims to place stringent constraints on the glitch model and nuclear forces. Meanwhile, with advancements in many-body theory and observational instruments, future efforts should focus on reducing model uncertainties and fostering multi-wavelength collaborations in glitch data analysis~\citep{2024MNRAS.533.4274L,2024arXiv240815022L}.
On one hand, non-relativistic and relativistic full quantum calculations, accounting for uncertainties in nuclear forces, can provide more accurate and detailed insights into pinning energies within the inner crust. Such calculations would be more realistic if one incorporates the effects of pasta structures at the base of the inner crust.
On the other hand, high-accuracy facilities like FAST, SKA, eXTP, with the help of multi-wavelength collaborative observations will offer a more unified understanding of stellar interiors and magnetospheres.
These advancements may lead to the development of a universal glitch model for Vela-like pulsars, offering a comprehensive description of internal superfluid dynamics. This synergy between theoretical and observational progress could greatly enhance our understanding of the nuclear force, particularly its poorly understood isospin dependence, as revealed in the present study.

\section*{Acknowledgments}
We are thankful to K. Sekizawa and the XMU neutron star group for helpful discussions.
The work is supported by National SKA Program of China (No.~2020SKA0120300), National Natural Science Foundation of China (grant Nos.~12273028, 12494572).

\bibliography{ref}

\begin{thebibliography}{}
\expandafter\ifx\csname natexlab\endcsname\relax\def\natexlab#1{#1}\fi
\providecommand{\url}[1]{\href{#1}{#1}}
\providecommand{\dodoi}[1]{doi:~\href{http://doi.org/#1}{\nolinkurl{#1}}}
\providecommand{\doeprint}[1]{\href{http://ascl.net/#1}{\nolinkurl{http://ascl.net/#1}}}
\providecommand{\doarXiv}[1]{\href{https://arxiv.org/abs/#1}{\nolinkurl{https://arxiv.org/abs/#1}}}

\bibitem[{Abbott {et~al.}(2017)Abbott, Abbott, Abbott, Acernese, Ackley, Adams,
  Adams, Addesso, Adhikari, Adya, Affeldt, Afrough, Agarwal, Agathos, Agatsuma,
  Aggarwal, Aguiar, Aiello, Ain, Ajith, Allen, Allen, Allocca, Altin, Amato,
  Ananyeva, Anderson, Anderson, Angelova, Antier, Appert, Arai, Araya, Areeda,
  Arnaud, Arun, Ascenzi, Ashton, Ast, Aston, Astone, Atallah, Aufmuth, Aulbert,
  AultONeal, Austin, Avila-Alvarez, Babak, Bacon, Bader, Bae, Bailes, Baker,
  Baldaccini, Ballardin, Ballmer, Banagiri, Barayoga, Barclay, Barish, Barker,
  Barkett, Barone, Barr, Barsotti, Barsuglia, Barta, Barthelmy, Bartlett,
  Bartos, Bassiri, Basti, Batch, Bawaj, Bayley, Bazzan, B{\'{e}}csy, Beer,
  Bejger, Belahcene, Bell, Berger, Bergmann, Bernuzzi, Bero, Berry, Bersanetti,
  Bertolini, Betzwieser, Bhagwat, Bhandare, Bilenko, Billingsley, Billman,
  Birch, Birney, Birnholtz, Biscans, Biscoveanu, Bisht, Bitossi, Biwer,
  Bizouard, Blackburn, Blackman, Blair, Blair, Blair, Bloemen, Bock, Bode,
  Boer, Bogaert, Bohe, Bondu, Bonilla, Bonnand, Boom, Bork, Boschi, Bose,
  Bossie, Bouffanais, Bozzi, Bradaschia, Brady, Branchesi, Brau, Briant,
  Brillet, Brinkmann, Brisson, Brockill, Broida, Brooks, Brown, Brown, Brunett,
  Buchanan, Buikema, Bulik, Bulten, Buonanno, Buskulic, Buy, Byer, Cabero,
  Cadonati, Cagnoli, Cahillane, Bustillo, Callister, Calloni, Camp, Canepa,
  Canizares, Cannon, Cao, Cao, Capano, Capocasa, Carbognani, Caride, Carney,
  Carullo, Diaz, Casentini, Caudill, Cavagli{\`{a}}, Cavalier, Cavalieri,
  Cella, Cepeda, Cerd{\'{a}}-Dur{\'{a}}n, Cerretani, Cesarini, Chamberlin,
  Chan, Chao, Charlton, Chase, Chassande-Mottin, Chatterjee, Chatziioannou,
  Cheeseboro, Chen, Chen, Chen, Cheng, Chia, Chincarini, Chiummo, Chmiel, Cho,
  Cho, Chow, Christensen, Chu, Chua, Chua, Chung, Chung, Ciani, Ciolfi,
  Cirelli, Cirone, Clara, Clark, Clearwater, Cleva, Cocchieri, Coccia, Cohadon,
  Cohen, Colla, Collette, Cominsky, Constancio, Conti, Cooper, Corban, Corbitt,
  Cordero-Carri{\'{o}}n, Corley, Cornish, Corsi, Cortese, Costa, Coughlin,
  Coughlin, Coulon, Countryman, Couvares, Covas, Cowan, Coward, Cowart, Coyne,
  Coyne, Creighton, Creighton, Cripe, Crowder, Cullen, Cumming, Cunningham,
  Cuoco, Canton, D{\'{a}}lya, Danilishin, D'Antonio, Danzmann, Dasgupta, Costa,
  Dattilo, Dave, Davier, Davis, Daw, Day, De, DeBra, Degallaix, Laurentis,
  Del{\'{e}}glise, Pozzo, Demos, Denker, Dent, Pietri, Dergachev, Rosa, DeRosa,
  Rossi, DeSalvo, de~Varona, Devenson, Dhurandhar, D{\'{\i}}az, Dietrich,
  Fiore, Giovanni, Girolamo, Lieto, Pace, Palma, Renzo, Doctor, Dolique,
  Donovan, Dooley, Doravari, Dorrington, Douglas, {\'{A}}lvarez, Downes, Drago,
  Dreissigacker, Driggers, Du, Ducrot, Dudi, Dupej, Dwyer, Edo, Edwards,
  Effler, Eggenstein, Ehrens, Eichholz, Eikenberry, Eisenstein, Essick,
  Estevez, Etienne, Etzel, Evans, Evans, Factourovich, Fafone, Fair, Fairhurst,
  Fan, Farinon, Farr, Farr, Fauchon-Jones, Favata, Fays, Fee, Fehrmann, Feicht,
  Fejer, Fernandez-Galiana, Ferrante, Ferreira, Ferrini, Fidecaro, Finstad,
  Fiori, Fiorucci, Fishbach, Fisher, Fitz-Axen, Flaminio, Fletcher, Fong, Font,
  Forsyth, Forsyth, Fournier, Frasca, Frasconi, Frei, Freise, Frey, Frey,
  Fries, Fritschel, Frolov, Fulda, Fyffe, Gabbard, Gadre, Gaebel, Gair,
  Gammaitoni, Ganija, Gaonkar, Garcia-Quiros, Garufi, Gateley, Gaudio, Gaur,
  Gayathri, Gehrels, Gemme, Genin, Gennai, George, George, Gergely, Germain,
  Ghonge, Ghosh, Ghosh, Ghosh, Giaime, Giardina, Giazotto, Gill, Glover, Goetz,
  Goetz, Gomes, Goncharov, Gonz{\'{a}}lez, Castro, Gopakumar, Gorodetsky,
  Gossan, Gosselin, Gouaty, Grado, Graef, Granata, Grant, Gras, Gray, Greco,
  Green, Gretarsson, Groot, Grote, Grunewald, Gruning, Guidi, Guo, Gupta,
  Gupta, Gushwa, Gustafson, Gustafson, Halim, Hall, Hall, Hamilton, Hammond,
  Haney, Hanke, Hanks, Hanna, Hannam, Hannuksela, Hanson, Hardwick, Harms,
  Harry, Harry, Hart, Haster, Haughian, Healy, Heidmann, Heintze, Heitmann,
  Hello, Hemming, Hendry, Heng, Hennig, Heptonstall, Heurs, Hild, Hinderer, Ho,
  Hoak, Hofman, Holt, Holz, Hopkins, Horst, Hough, Houston, Howell, Hreibi, Hu,
  Huerta, Huet, Hughey, Husa, Huttner, Huynh-Dinh, Indik, Inta, Intini, Isa,
  Isac, Isi, Iyer, Izumi, Jacqmin, Jani, Jaranowski, Jawahar,
  Jim{\'{e}}nez-Forteza, Johnson, Johnson-McDaniel, Jones, Jones, Jonker, Ju,
  Junker, Kalaghatgi, Kalogera, Kamai, Kandhasamy, Kang, Kanner, Kapadia,
  Karki, Karvinen, Kasprzack, Kastaun, Katolik, Katsavounidis, Katzman, Kaufer,
  Kawabe, K{\'{e}}f{\'{e}}lian, Keitel, Kemball, Kennedy, Kent, Key, Khalili,
  Khan, Khan, Khan, Khazanov, Kijbunchoo, Kim, Kim, Kim, Kim, Kim, Kim,
  Kimbrell, King, King, Kinley-Hanlon, Kirchhoff, Kissel, Kleybolte, Klimenko,
  Knowles, Koch, Koehlenbeck, Koley, Kondrashov, Kontos, Korobko, Korth,
  Kowalska, Kozak, Krämer, Kringel, Krishnan, Kr{\'{o}}lak, Kuehn, Kumar,
  Kumar, Kumar, Kuo, Kutynia, Kwang, Lackey, Lai, Landry, Lang, Lange, Lantz,
  Lanza, Larson, Lartaux-Vollard, Lasky, Laxen, Lazzarini, Lazzaro, Leaci,
  Leavey, Lee, Lee, Lee, Lee, Lee, Lehmann, Lenon, Leon, Leonardi, Leroy,
  Letendre, Levin, Li, Linker, Littenberg, Liu, Liu, Lo, Lockerbie, London,
  Lord, Lorenzini, Loriette, Lormand, Losurdo, Lough, Lousto, Lovelace, Lück,
  Lumaca, Lundgren, Lynch, Ma, Macas, Macfoy, Machenschalk, MacInnis, Macleod,
  Hernandez, Maga{\~{n}}a-Sandoval, Zertuche, Magee, Majorana, Maksimovic, Man,
  Mandic, Mangano, Mansell, Manske, Mantovani, Marchesoni, Marion, M{\'{a}}rka,
  M{\'{a}}rka, Markakis, Markosyan, Markowitz, Maros, Marquina, Marsh,
  Martelli, Martellini, Martin, Martin, Martynov, Marx, Mason, Massera,
  Masserot, Massinger, Masso-Reid, Mastrogiovanni, Matas, Matichard, Matone,
  Mavalvala, Mazumder, McCarthy, McClelland, McCormick, McCuller, McGuire,
  McIntyre, McIver, McManus, McNeill, McRae, McWilliams, Meacher, Meadors,
  Mehmet, Meidam, Mejuto-Villa, Melatos, Mendell, Mercer, Merilh, Merzougui,
  Meshkov, Messenger, Messick, Metzdorff, Meyers, Miao, Michel, Middleton,
  Mikhailov, Milano, Miller, Miller, Miller, Millhouse, Milovich-Goff,
  Minazzoli, Minenkov, Ming, Mishra, Mitra, Mitrofanov, Mitselmakher,
  Mittleman, Moffa, Moggi, Mogushi, Mohan, Mohapatra, Molina, Montani, Moore,
  Moraru, Moreno, Morisaki, Morriss, Mours, Mow-Lowry, Mueller, Muir,
  Mukherjee, Mukherjee, Mukherjee, Mukund, Mullavey, Munch, Mu{\~{n}}iz,
  Muratore, Murray, Nagar, Napier, Nardecchia, Naticchioni, Nayak, Neilson,
  Nelemans, Nelson, Nery, Neunzert, Nevin, Newport, Newton, Ng, Nguyen, Nguyen,
  Nichols, Nielsen, Nissanke, Nitz, Noack, Nocera, Nolting, North, Nuttall,
  Oberling, O'Dea, Ogin, Oh, Oh, Ohme, Okada, Oliver, Oppermann, Oram,
  O'Reilly, Ormiston, Ortega, O'Shaughnessy, Ossokine, Ottaway, Overmier, Owen,
  Pace, Page, Page, Pai, Pai, Palamos, Palashov, Palomba, Pal-Singh, Pan, Pan,
  Pang, Pang, Pankow, Pannarale, Pant, Paoletti, Paoli, Papa, Parida, Parker,
  Pascucci, Pasqualetti, Passaquieti, Passuello, Patil, Patricelli, Pearlstone,
  Pedraza, Pedurand, Pekowsky, Pele, Penn, Perez, Perreca, Perri, Pfeiffer,
  Phelps, Piccinni, Pichot, Piergiovanni, Pierro, Pillant, Pinard, Pinto,
  Pirello, Pitkin, Poe, Poggiani, Popolizio, Porter, Post, Powell, Prasad,
  Pratt, Pratten, Predoi, Prestegard, Prijatelj, Principe, Privitera, Prix,
  Prodi, Prokhorov, Puncken, Punturo, Puppo, Pürrer, Qi, Quetschke, Quintero,
  Quitzow-James, Raab, Rabeling, Radkins, Raffai, Raja, Rajan, Rajbhandari,
  Rakhmanov, Ramirez, Ramos-Buades, Rapagnani, Raymond, Razzano, Read,
  Regimbau, Rei, Reid, Reitze, Ren, Reyes, Ricci, Ricker, Rieger, Riles, Rizzo,
  Robertson, Robie, Robinet, Rocchi, Rolland, Rollins, Roma, Romano, Romano,
  Romel, Romie, Rosi{\'{n}}ska, Ross, Rowan, Rüdiger, Ruggi, Rutins, Ryan,
  Sachdev, Sadecki, Sadeghian, Sakellariadou, Salconi, Saleem, Salemi,
  Samajdar, Sammut, Sampson, Sanchez, Sanchez, Sanchis-Gual, Sandberg, Sanders,
  Sassolas, Sathyaprakash, Saulson, Sauter, Savage, Sawadsky, Schale, Scheel,
  Scheuer, Schmidt, Schmidt, Schnabel, Schofield, Schönbeck, Schreiber,
  Schuette, Schulte, Schutz, Schwalbe, Scott, Scott, Seidel, Sellers, Sengupta,
  Sentenac, Sequino, Sergeev, Shaddock, Shaffer, Shah, Shahriar, Shaner, Shao,
  Shapiro, Shawhan, Sheperd, Shoemaker, Shoemaker, Siellez, Siemens,
  Sieniawska, Sigg, Silva, Singer, Singh, Singhal, Sintes, Slagmolen, Smith,
  Smith, Smith, Somala, Son, Sonnenberg, Sorazu, Sorrentino, Souradeep,
  Spencer, Srivastava, Staats, Staley, Steinke, Steinlechner, Steinlechner,
  Steinmeyer, Stevenson, Stone, Stops, Strain, Stratta, Strigin, Strunk,
  Sturani, Stuver, Summerscales, Sun, Sunil, Suresh, Sutton, Swinkels,
  Szczepa{\'{n}}czyk, Tacca, Tait, Talbot, Talukder, Tanner, T{\'{a}}pai,
  Taracchini, Tasson, Taylor, Taylor, Tewari, Theeg, Thies, Thomas, Thomas,
  Thomas, Thorne, Thorne, Thrane, Tiwari, Tiwari, Tokmakov, Toland, Tonelli,
  Tornasi, Torres-Forn{\'{e}}, Torrie, Töyrä, Travasso, Traylor, Trinastic,
  Tringali, Trozzo, Tsang, Tse, Tso, Tsukada, Tsuna, Tuyenbayev, Ueno, Ugolini,
  Unnikrishnan, Urban, Usman, Vahlbruch, Vajente, Valdes, Vallisneri, van
  Bakel, van Beuzekom, van~den Brand, Broeck, Vander-Hyde, van~der Schaaf, van
  Heijningen, van Veggel, Vardaro, Varma, Vass, Vas{\'{u}}th, Vecchio,
  Vedovato, Veitch, Veitch, Venkateswara, Venugopalan, Verkindt, Vetrano,
  Vicer{\'{e}}, Viets, Vinciguerra, Vine, Vinet, Vitale, Vo, Vocca, Vorvick,
  Vyatchanin, Wade, Wade, Wade, Walet, Walker, Wallace, Walsh, Wang, Wang,
  Wang, Wang, Wang, Ward, Warner, Was, Watchi, Weaver, Wei, Weinert, Weinstein,
  Weiss, Wen, Wessel, We{\ss}els, Westerweck, Westphal, Wette, Whelan,
  Whitcomb, Whiting, Whittle, Wilken, Williams, Williams, Williamson, Willis,
  Willke, Wimmer, Winkler, Wipf, Wittel, Woan, Woehler, Wofford, Wong, Worden,
  Wright, Wu, Wysocki, Xiao, Yamamoto, Yancey, Yang, Yap, Yazback, Yu, Yu,
  Yvert, Zadro{\.{z}}ny, Zanolin, Zelenova, Zendri, Zevin, Zhang, Zhang, Zhang,
  Zhang, Zhao, Zhou, Zhou, Zhu, Zhu, Zimmerman, Zucker, \&
  and}]{Abbott2017_PRL119-161101}
Abbott, B., Abbott, R., Abbott, T., {et~al.} 2017, Phys. Rev. Lett, 119,
  161101, \dodoi{10.1103/physrevlett.119.161101}

\bibitem[{Ala\~na {et~al.}(2024)Ala\~na, Modugno, Capuzzi, \&
  Jezek}]{Alana2024_PRA110-023306}
Ala\~na, A., Modugno, M., Capuzzi, P., \& Jezek, D.~M. 2024, Phys. Rev. A, 110,
  023306, \dodoi{10.1103/PhysRevA.110.023306}

\bibitem[{Alpar(1977)}]{Alpar1977_ApJ213-527}
Alpar, M.~A. 1977, Astrophys. J, 213, 527, \dodoi{10.1086/155183}

\bibitem[{Alpar {et~al.}(1993)Alpar, Chau, Cheng, \&
  Pines}]{Alpar1993_ApJ409-345}
Alpar, M.~A., Chau, H.~F., Cheng, K.~S., \& Pines, D. 1993, Astrophys. J, 409,
  345, \dodoi{10.1086/172668}

\bibitem[{Alpar {et~al.}(1984)Alpar, Pines, Anderson, \&
  Shaham}]{Alpar1984_ApJ276-325}
Alpar, M.~A., Pines, D., Anderson, P.~W., \& Shaham, J. 1984, Astrophys. J,
  276, 325, \dodoi{10.1086/161616}

\bibitem[{Anderson \& ITOH(1975)}]{Anderson1975_Nature256-25}
Anderson, P.~W., \& ITOH, N. 1975, Nature, 256, 25, \dodoi{10.1038/256025a0}

\bibitem[{Antonelli {et~al.}(2022)Antonelli, Montoli, \&
  Pizzochero}]{Antonelli2022_NSPhysicsGlitches}
Antonelli, M., Montoli, A., \& Pizzochero, P.~M. 2022, Insights Into the
  Physics of Neutron Star Interiors from Pulsar Glitches (WORLD SCIENTIFIC),
  219--281, \dodoi{10.1142/9789811220944_0007}

\bibitem[{Antonopoulou {et~al.}(2022)Antonopoulou, Haskell, \&
  Espinoza}]{Antonopoulou2022_RPP85-126901}
Antonopoulou, D., Haskell, B., \& Espinoza, C.~M. 2022, Rep. Prog. Phys, 85,
  126901, \dodoi{10.1088/1361-6633/ac9ced}

\bibitem[{Ashton {et~al.}(2019)Ashton, Lasky, Graber, \&
  Palfreyman}]{Ashton2019_NA36-844}
Ashton, G., Lasky, P.~D., Graber, V., \& Palfreyman, J. 2019, Nature Astronomy,
  36, 844, \dodoi{10.1038/s41550-019-0844-6}

\bibitem[{Avogadro {et~al.}(2008)Avogadro, Barranco, Broglia, \&
  Vigezzi}]{Avogadro2008_NPA811-378}
Avogadro, P., Barranco, F., Broglia, R., \& Vigezzi, E. 2008, Nucl. Phys. A,
  811, 378, \dodoi{10.1016/j.nuclphysa.2008.07.010}

\bibitem[{Avogadro {et~al.}(2007)Avogadro, Barranco, Broglia, \&
  Vigezzi}]{Avogadro2007_PRC75-012805}
Avogadro, P., Barranco, F., Broglia, R.~A., \& Vigezzi, E. 2007, Phys. Rev. C,
  75, 012805, \dodoi{10.1103/PhysRevC.75.012805}

\bibitem[{Balberg \& Barnea(1998)}]{Balberg1998_PRC57-409}
Balberg, S., \& Barnea, N. 1998, Phys. Rev. C, 57, 409,
  \dodoi{10.1103/PhysRevC.57.409}

\bibitem[{Basu {et~al.}(2021)Basu, Shaw, Antonopoulou, Keith, Lyne, Mickaliger,
  Stappers, Weltevrede, \& Jordan}]{Basu2021_MNRAS510-4049}
Basu, A., Shaw, B., Antonopoulou, D., {et~al.} 2021, Mon. Not. R. Astron. Soc,
  510, 4049, \dodoi{10.1093/mnras/stab3336}

\bibitem[{Baym {et~al.}(1969)Baym, Pethick, Pines, \&
  Ruderman}]{Baym1969_Nature224-872}
Baym, G., Pethick, C., Pines, D., \& Ruderman, M. 1969, Nature, 224, 872,
  \dodoi{10.1038/224872a0}

\bibitem[{Bender {et~al.}(2003)Bender, Heenen, \&
  Reinhard}]{Bender2003_RMP75-121}
Bender, M., Heenen, P.-H., \& Reinhard, P.-G. 2003, Rev. Mod. Phys, 75, 121,
  \dodoi{10.1103/revmodphys.75.121}

\bibitem[{Bland {et~al.}(2024)Bland, Ferlaino, Mannarelli, Poli, \&
  Trabucco}]{Bland2024_FBS65-81}
Bland, T., Ferlaino, F., Mannarelli, M., Poli, E., \& Trabucco, S. 2024,
  Few-Body System, 65, \dodoi{10.1007/s00601-024-01949-7}

\bibitem[{Boguta \& Bodmer(1977)}]{Boguta1977_NPA292-413}
Boguta, J., \& Bodmer, A. 1977, Nucl. Phys. A, 292, 413,
  \dodoi{10.1016/0375-9474(77)90626-1}

\bibitem[{Campbell(1979)}]{Campbell1979_PRL43-1336}
Campbell, L.~J. 1979, Phys. Rev. Lett, 43, 1336,
  \dodoi{10.1103/PhysRevLett.43.1336}

\bibitem[{Chamel(2013)}]{Chamel2013_PRL110-011101}
Chamel, N. 2013, Phys. Rev. Lett., 110, 011101,
  \dodoi{10.1103/PhysRevLett.110.011101}

\bibitem[{Chamel(2024)}]{Chamel2024_arXiv2412.05599}
---. 2024.
\newblock \doarXiv{2412.05599}

\bibitem[{Donati \& Pizzochero(2004)}]{Donati2004_NPA742-363}
Donati, P., \& Pizzochero, P.~M. 2004, Nucl. Phys. A, 742, 363,
  \dodoi{https://doi.org/10.1016/j.nuclphysa.2004.07.002}

\bibitem[{Donati \& Pizzochero(2006)}]{Donati2006_PLB640-74}
---. 2006, Phys. Lett. B, 640, 74, \dodoi{10.1016/j.physletb.2006.07.047}

\bibitem[{Dutra {et~al.}(2014)Dutra, Lourenco, Avancini, Carlson, Delfino,
  Menezes, Provid\^{e}ncia, Typel, \& Stone}]{Dutra2014_PRC90-055203}
Dutra, M., Lourenco, O., Avancini, S.~S., {et~al.} 2014, Phys. Rev. C, 90,
  055203, \dodoi{10.1103/PhysRevC.90.055203}

\bibitem[{Epstein \& Baym(1988)}]{Epstein1988_ApJ328-680}
Epstein, R.~I., \& Baym, G. 1988, Astrophys. J, 328, 680,
  \dodoi{10.1086/166325}

\bibitem[{Espinoza {et~al.}(2011)Espinoza, Lyne, Stappers, \&
  Kramer}]{Espinoza2011_MNRAS414-1679}
Espinoza, C.~M., Lyne, A.~G., Stappers, B.~W., \& Kramer, M. 2011, Mon. Not. R.
  Astron. Soc, 414, 1679, \dodoi{10.1111/j.1365-2966.2011.18503.x}

\bibitem[{Fantina {et~al.}(2016)Fantina, Chamel, Mutafchieva, Stoyanov,
  Mihailov, \& Pavlov}]{Fantina2016_PRC93-015801}
Fantina, A.~F., Chamel, N., Mutafchieva, Y.~D., {et~al.} 2016, Phys. Rev. C,
  93, 015801, \dodoi{10.1103/physrevc.93.015801}

\bibitem[{Fetter {et~al.}(1971)Fetter, Walecka, \&
  Kadanoff}]{Fetter1971_PT25-54}
Fetter, A.~L., Walecka, J.~D., \& Kadanoff, L.~P. 1971, Physics Today, 25, 54.
\newblock \url{https://api.semanticscholar.org/CorpusID:5661190}

\bibitem[{Fuchs {et~al.}(1995)Fuchs, Lenske, \& Wolter}]{Fuchs1995_PRC52-3043}
Fuchs, C., Lenske, H., \& Wolter, H.~H. 1995, Phys. Rev. C, 52, 3043,
  \dodoi{10.1103/PhysRevC.52.3043}

\bibitem[{Glendenning(1996)}]{Glendenning1996_CompactStars}
Glendenning, N.~K. 1996, Compact Stars: Nuclear Physics, Particle Physics and
  General Relativity (Springer US)

\bibitem[{Graber {et~al.}(2018)Graber, Cumming, \&
  Andersson}]{Graber2018_ApJ865-23}
Graber, V., Cumming, A., \& Andersson, N. 2018, Astrophys. J, 865, 23,
  \dodoi{10.3847/1538-4357/aad776}

\bibitem[{Grill \& Pizzochero(2012)}]{Grill2012_JPCS342-012004}
Grill, F., \& Pizzochero, P. 2012, Jour. Phys. Conf. Ser, 342, 012004,
  \dodoi{10.1088/1742-6596/342/1/012004}

\bibitem[{Gügercinoğlu \& Alpar(2020)}]{Guegercinoglu2020_MNRAS496-2506}
Gügercinoğlu, E., \& Alpar, M.~A. 2020, Mon. Not. R. Astron. Soc, 496, 2506,
  \dodoi{10.1093/mnras/staa1672}

\bibitem[{Haskell \& Melatos(2015)}]{Haskell2015_IJMPD24-1530008}
Haskell, B., \& Melatos, A. 2015, Inter. Jour. Mod. Phys. D, 24, 1530008,
  \dodoi{10.1142/s0218271815300086}

\bibitem[{Haskell {et~al.}(2013)Haskell, Pizzochero, \&
  Seveso}]{Haskell2013_ApJ764-L25}
Haskell, B., Pizzochero, P.~M., \& Seveso, S. 2013, Astrophys. J, 764, L25,
  \dodoi{10.1088/2041-8205/764/2/l25}

\bibitem[{Haskell {et~al.}(2011)Haskell, Pizzochero, \&
  Sidery}]{Haskell2011_MNRAS420-658}
Haskell, B., Pizzochero, P.~M., \& Sidery, T. 2011, Mon. Not. R. Astron. Soc,
  420, 658, \dodoi{10.1111/j.1365-2966.2011.20080.x}

\bibitem[{Hooker {et~al.}(2013)Hooker, Newton, \&
  Li}]{Hooker2013_JPCS420-012153}
Hooker, J., Newton, W.~G., \& Li, B.-A. 2013, Jour. Phys. Conf. Ser, 420,
  012153, \dodoi{10.1088/1742-6596/420/1/012153}

\bibitem[{Khodel {et~al.}(1996)Khodel, Khodel, \&
  Clark}]{Khodel1996_NPA598-390}
Khodel, V., Khodel, V., \& Clark, J. 1996, Nucl. Phys. A, 598, 390,
  \dodoi{https://doi.org/10.1016/0375-9474(95)00477-7}

\bibitem[{Klausner {et~al.}(2023)Klausner, Barranco, Pizzochero, Roca-Maza, \&
  Vigezzi}]{Klausner2023_PRC108-035808}
Klausner, P., Barranco, F., Pizzochero, P.~M., Roca-Maza, X., \& Vigezzi, E.
  2023, Phys. Rev. C, 108, 035808, \dodoi{10.1103/PhysRevC.108.035808}

\bibitem[{Lalazissis {et~al.}(2005)Lalazissis, Niksi{\'{c}}, Vretenar, \&
  Ring}]{Lalazissis2005_PRC71-024312}
Lalazissis, G.~A., Niksi{\'{c}}, T., Vretenar, D., \& Ring, P. 2005, Phys. Rev.
  C, 71, 024312, \dodoi{10.1103/PhysRevC.71.024312}

\bibitem[{Li {et~al.}(2016)Li, Dong, Wang, \& Xu}]{Li2016_ApJSupp223-16}
Li, A., Dong, J.~M., Wang, J.~B., \& Xu, R.~X. 2016, Astrophys. J. Suppl. Ser.,
  223, 16, \dodoi{10.3847/0067-0049/223/1/16}

\bibitem[{{Li} {et~al.}(2020){Li}, {Zhu}, {Zhou}, {Dong}, {Hu}, \&
  {Xia}}]{2020JHEAp..28...19L}
{Li}, A., {Zhu}, Z.~Y., {Zhou}, E.~P., {et~al.} 2020, Journal of High Energy
  Astrophysics, 28, 19, \dodoi{10.1016/j.jheap.2020.07.001}

\bibitem[{Li \& Sedrakian(2019)}]{Li2019_PRC100-015809}
Li, J.~J., \& Sedrakian, A. 2019, Phys. Rev. C, 100, 015809,
  \dodoi{10.1103/PhysRevC.100.015809}

\bibitem[{{Liu} {et~al.}(2024{\natexlab{a}}){Liu}, {Yuan}, {Ge}, {Ye}, {Zhou},
  {Dang}, {Zhou}, {G{\"u}gercino{\u{g}}lu}, {Wang}, {Wang}, {Li}, {Li}, \&
  {Wang}}]{2024MNRAS.533.4274L}
{Liu}, P., {Yuan}, J.~P., {Ge}, M.~Y., {et~al.} 2024{\natexlab{a}}, \mnras,
  533, 4274, \dodoi{10.1093/mnras/stae1973}

\bibitem[{{Liu} {et~al.}(2024{\natexlab{b}}){Liu}, {Yuan}, {Ge}, {Ye}, {Zhou},
  {Dang}, {Zhou}, {G{\"u}gercino{\u{g}}lu}, {Tu}, {Wang}, {Li}, {Li}, \&
  {Wang}}]{2024arXiv240815022L}
---. 2024{\natexlab{b}}, arXiv e-prints, arXiv:2408.15022,
  \dodoi{10.48550/arXiv.2408.15022}

\bibitem[{Lombardo \& Schulze(2000)}]{Lombardo2000_arXiv0012209}
Lombardo, U., \& Schulze, H.~J. 2000, Lect.Notes Phys.578:30-53,2001.
\newblock \doarXiv{astro-ph/0012209}

\bibitem[{Long {et~al.}(2004)Long, Meng, Giai, \& Zhou}]{Long2004_PRC69-034319}
Long, W., Meng, J., Giai, N.~V., \& Zhou, S.-G. 2004, Phys. Rev. C, 69, 034319,
  \dodoi{10.1103/PhysRevC.69.034319}

\bibitem[{Lower {et~al.}(2021)Lower, Johnston, Dunn, Shannon, Bailes, Dai,
  Kerr, Manchester, Melatos, Oswald, Parthasarathy, Sobey, \&
  Weltevrede}]{Lower2021_MNRAS508-3251}
Lower, M.~E., Johnston, S., Dunn, L., {et~al.} 2021, Mon. Not. R. Astron. Soc,
  508, 3251, \dodoi{10.1093/mnras/stab2678}

\bibitem[{Melatos \& Millhouse(2023)}]{Melatos2023_ApJ948-106}
Melatos, A., \& Millhouse, M. 2023, Astrophys. J, 948, 106,
  \dodoi{10.3847/1538-4357/acbb6e}

\bibitem[{Negele \& Vautherin(1973)}]{Negele1973_NPA207-298}
Negele, J., \& Vautherin, D. 1973, Nuc. Phys. A, 207, 298,
  \dodoi{10.1016/0375-9474(73)90349-7}

\bibitem[{Neill {et~al.}(2024)Neill, Tsang, \& Newton}]{Neill2024_MNRAS532-827}
Neill, D., Tsang, D., \& Newton, W.~G. 2024, Mon. Not. R. Astron. Soc, 532,
  827, \dodoi{10.1093/mnras/stae1481}

\bibitem[{Newton {et~al.}(2013)Newton, Murphy, Hooker, \&
  Li}]{Newton2013_ApJL779-L4}
Newton, W.~G., Murphy, K., Hooker, J., \& Li, B.-A. 2013, Astrophys. J. Lett,
  779, L4, \dodoi{10.1088/2041-8205/779/1/l4}

\bibitem[{Nikšić {et~al.}(2011)Nikšić, Vretenar, \&
  Ring}]{Niksic2011_PPNP66-519}
Nikšić, T., Vretenar, D., \& Ring, P. 2011, Prog. Part. Nucl. Phys., 66, 519,
  \dodoi{10.1016/j.ppnp.2011.01.055}

\bibitem[{Oertel {et~al.}(2017)Oertel, Hempel, Kl\"ahn, \&
  Typel}]{Oertel2017_RMP89-015007}
Oertel, M., Hempel, M., Kl\"ahn, T., \& Typel, S. 2017, Rev. Mod. Phys, 89,
  015007, \dodoi{10.1103/RevModPhys.89.015007}

\bibitem[{Oppenheimer \& Volkoff(1939)}]{Oppenheimer1939_PR055-374}
Oppenheimer, J.~R., \& Volkoff, G.~M. 1939, Phys. Rev, 055, 374,
  \dodoi{10.1103/PhysRev.55.374}

\bibitem[{Oyamatsu \& Iida(2007)}]{Oyamatsu2007_PRC75-015801}
Oyamatsu, K., \& Iida, K. 2007, Phys. Rev. C, 75, 015801,
  \dodoi{10.1103/physrevc.75.015801}

\bibitem[{Pizzochero(2011)}]{Pizzochero2011_ApJ743-L20}
Pizzochero, P.~M. 2011, Astrophys. J., 743, L20,
  \dodoi{10.1088/2041-8205/743/1/l20}

\bibitem[{Pizzochero {et~al.}(1997)Pizzochero, Viverit, \&
  Broglia}]{Pizzochero1997_PRL79-3347}
Pizzochero, P.~M., Viverit, L., \& Broglia, R.~A. 1997, Phys. Rev. Lett., 79,
  3347, \dodoi{10.1103/PhysRevLett.79.3347}

\bibitem[{{Pizzochero, P. M.} {et~al.}(2020){Pizzochero, P. M.}, {Montoli, A.},
  \& {Antonelli, M.}}]{Pizzochero2020_A&A636-A101}
{Pizzochero, P. M.}, {Montoli, A.}, \& {Antonelli, M.} 2020, A\&A, 636, A101,
  \dodoi{10.1051/0004-6361/201937019}

\bibitem[{Poli {et~al.}(2023)Poli, Bland, White, Mark, Ferlaino, Trabucco, \&
  Mannarelli}]{Poli2023_PRL131-223401}
Poli, E., Bland, T., White, S.~J., {et~al.} 2023, Phys. Rev. Lett, 131, 223401,
  \dodoi{10.1103/physrevlett.131.223401}

\bibitem[{Reinhard(1989)}]{Reinhard1989_RPP52-439}
Reinhard, P.~G. 1989, Rep. Prog. Phys., 52, 439,
  \dodoi{10.1088/0034-4885/52/4/002}

\bibitem[{Riley {et~al.}(2021)Riley, Watts, Ray, Bogdanov, Guillot, Morsink,
  Bilous, Arzoumanian, Choudhury, Deneva, Gendreau, Harding, Ho, Lattimer,
  Loewenstein, Ludlam, Markwardt, Okajima, Prescod-Weinstein, Remillard, Wolff,
  Fonseca, Cromartie, Kerr, Pennucci, Parthasarathy, Ransom, Stairs, Guillemot,
  \& Cognard}]{Riley2021_ApJL918-L27}
Riley, T.~E., Watts, A.~L., Ray, P.~S., {et~al.} 2021, Astrophys. J. Lett, 918,
  L27, \dodoi{10.3847/2041-8213/ac0a81}

\bibitem[{Ring(1996)}]{Ring1996_PPNP37-193}
Ring, P. 1996, Prog. Part. Nucl. Phys., 37, 193,
  \dodoi{10.1016/0146-6410(96)00054-3}

\bibitem[{Rong {et~al.}(2021)Rong, Tu, \& Zhou}]{Rong2021_PRC104-054321}
Rong, Y.-T., Tu, Z.-H., \& Zhou, S.-G. 2021, Phys. Rev. C, 104, 054321,
  \dodoi{10.1103/physrevc.104.054321}

\bibitem[{Rong {et~al.}(2020)Rong, Zhao, \& Zhou}]{Rong2020_PLB807-135533}
Rong, Y.-T., Zhao, P., \& Zhou, S.-G. 2020, Phys. Lett. B, 807, 135533,
  \dodoi{10.1016/j.physletb.2020.135533}

\bibitem[{Schaffner-Bielich \& Gal(2000)}]{Schaffner-Bielich2000_PRC62-034311}
Schaffner-Bielich, J., \& Gal, A. 2000, Phys. Rev. C, 62, 034311,
  \dodoi{10.1103/PhysRevC.62.034311}

\bibitem[{Serot(1992)}]{Serot1992_RPP55-1855}
Serot, B.~D. 1992, Rept. Prog. Phys", 55, 1855,
  \dodoi{10.1088/0034-4885/55/11/001}

\bibitem[{Seveso {et~al.}(2016)Seveso, Pizzochero, Grill, \&
  Haskell}]{Seveso2016_MNRAS455-3952}
Seveso, S., Pizzochero, P.~M., Grill, F., \& Haskell, B. 2016, Mon. Not. R.
  Astron. Soc, 455, 3952, \dodoi{10.1093/mnras/stv2579}

\bibitem[{Seveso {et~al.}(2012)Seveso, Pizzochero, \&
  Haskell}]{Seveso2012_MNRAS427-1089}
Seveso, S., Pizzochero, P.~M., \& Haskell, B. 2012, Mon. Not. R. Astron. Soc,
  427, 1089, \dodoi{10.1111/j.1365-2966.2012.21906.x}

\bibitem[{Shang \& Li(2021)}]{Shang2021_ApJ923-108}
Shang, X., \& Li, A. 2021, Astrophys. J, 923, 108,
  \dodoi{10.3847/1538-4357/ac2e94}

\bibitem[{Shen(2002)}]{Shen2002_PRC65-035802}
Shen, H. 2002, Phys. Rev. C, 65, 035802, \dodoi{10.1103/PhysRevC.65.035802}

\bibitem[{Sun {et~al.}(2023)Sun, Miao, Sun, \& Li}]{Sun2023_ApJ942-55}
Sun, X., Miao, Z., Sun, B., \& Li, A. 2023, Astrophys. J, 942, 55,
  \dodoi{10.3847/1538-4357/ac9d9a}

\bibitem[{Tian {et~al.}(2009)Tian, Ma, \& Ring}]{Tian2009_PLB676-44}
Tian, Y., Ma, Z.~Y., \& Ring, P. 2009, Phys. Lett. B, 676, 44,
  \dodoi{10.1016/j.physletb.2009.04.067}

\bibitem[{Tolman(1939)}]{Tolman1939_PR055-364}
Tolman, R.~C. 1939, Phys. Rev., 55, 364, \dodoi{10.1103/PhysRev.55.364}

\bibitem[{Tsakadze \& Tsakadze(1980)}]{Tsakadze1980_JLTP39-649}
Tsakadze, J.~S., \& Tsakadze, S.~J. 1980, J. Low. Temp. Phys, 39, 649,
  \dodoi{10.1007/bf00114899}

\bibitem[{Tu \& Zhou(2022)}]{Tu2022_ApJ925-16}
Tu, Z.-H., \& Zhou, S.-G. 2022, Astrophys. J, 925, 16,
  \dodoi{10.3847/1538-4357/ac3996}

\bibitem[{Typel \& Wolter(1999)}]{Typel1999_NPA656-331}
Typel, S., \& Wolter, H. 1999, Nucl. Phys. A, 656, 331,
  \dodoi{10.1016/S0375-9474(99)00310-3}

\bibitem[{Vinciguerra {et~al.}(2024)Vinciguerra, Salmi, Watts, Choudhury,
  Riley, Ray, Bogdanov, Kini, Guillot, Chakrabarty, Ho, Huppenkothen, Morsink,
  Wadiasingh, \& Wolff}]{Vinciguerra2024_ApJ961-62}
Vinciguerra, S., Salmi, T., Watts, A.~L., {et~al.} 2024, Astrophys. J, 961, 62,
  \dodoi{10.3847/1538-4357/acfb83}

\bibitem[{Walecka(1974)}]{Walecka1974_APNY83-491}
Walecka, J. 1974, Ann. Phys, 83, 491, \dodoi{10.1016/0003-4916(74)90208-5}

\bibitem[{Wang \& Shen(2010)}]{Wang2010_PRC81-025801}
Wang, Y.~N., \& Shen, H. 2010, Phys. Rev. C, 81, 025801,
  \dodoi{10.1103/PhysRevC.81.025801}

\bibitem[{Wu {et~al.}(2021)Wu, Bao, Shen, \& Xu}]{Wu2021_PRC104-015802}
Wu, X., Bao, S., Shen, H., \& Xu, R. 2021, Phys. Rev. C, 104, 015802,
  \dodoi{10.1103/PhysRevC.104.015802}

\bibitem[{Yan(2019)}]{Yan2019_RAA19-072}
Yan, Y. 2019, Res. Astron. Astrophys, 19, 072,
  \dodoi{10.1088/1674-4527/19/5/72}

\bibitem[{Yu {et~al.}(2012)Yu, Manchester, Hobbs, Johnston, Kaspi, Keith, Lyne,
  Qiao, Ravi, Sarkissian, Shannon, \& Xu}]{Yu2012_MNRAS429-688}
Yu, M., Manchester, R.~N., Hobbs, G., {et~al.} 2012, Mon. Not. R. Astron. Soc,
  429, 688, \dodoi{10.1093/mnras/sts366}

\bibitem[{Zhou {et~al.}(2022)Zhou, Gügercinoğlu, Yuan, Ge, \&
  Yu}]{Zhou2022_Universe8-641}
Zhou, S., Gügercinoğlu, E., Yuan, J., Ge, M., \& Yu, C. 2022, Universe, 8,
  641, \dodoi{10.3390/universe8120641}

\bibitem[{{Zhu} {et~al.}(2023){Zhu}, {Li}, {Hu}, \&
  {Shen}}]{2023PhRvC.108b5809Z}
{Zhu}, Z., {Li}, A., {Hu}, J., \& {Shen}, H. 2023, \prc, 108, 025809,
  \dodoi{10.1103/PhysRevC.108.025809}

\bibitem[{Zubieta {et~al.}(2024)Zubieta, García, del Palacio, Araujo~Furlan,
  Gancio, Lousto, Combi, \& Espinoza}]{Zubieta2024_AA689-A191}
Zubieta, E., García, F., del Palacio, S., {et~al.} 2024, Astron. Astrophys,
  689, A191, \dodoi{10.1051/0004-6361/202450441}

\end{thebibliography}
\bibliographystyle{aasjournal}

\end{document}